\documentclass{svjour3}                     % onecolumn (standard format)
\smartqed  % flush right qed marks, e.g. at end of proof
\usepackage{graphicx}
\begin{document}

\title{Modeling of complex oxide materials from the first principles:
systematic
applications to vanadates $R$VO$_3$ with distorted perovskite structure
\thanks{This work is partly supported by Grant-in-Aid for Scientific
Research (C) No. 20540337
from the
Ministry of Education, Culture, Sport, Science and Technology of
Japan.}
}

\author{Igor Solovyev
}

\institute{I. Solovyev \at
              National Institute for Materials Science, 1-2-1 Sengen, Tsukuba, Ibaraki 305-0047, Japan \\
              Tel.: +81-29-859-2619 \\
              Fax:  +81-29-859-2601 \\
              \email{SOLOVYEV.Igor@nims.go.jp}
}

\date{Received: date / Accepted: date}

\maketitle

\begin{abstract}
``Realistic modeling'' is a new direction of electronic structure calculations,
where the main emphasis is made on the construction of some effective low-energy model
entirely within a first-principle framework. Ideally, it is a model in form, but with all the
parameters derived rigorously, on the basis of first-principles electronic
structure calculations. The method is especially suit for transition-metal oxides and
other strongly correlated systems, whose electronic and magnetic
properties are predetermined by the behavior of some limited number of states located near
the Fermi level. After reviewing general ideas of realistic modeling,
we will illustrate abilities of this approach on the wide series of vanadates
$R$VO$_3$ ($R$$=$ La, Ce, Pr, Nd, Sm, Gd, Tb, Yb, and Y)
with distorted perovskite structure. Particular attention will be paid
to computational tools, which can be used for microscopic analysis
of different spin
and orbital states
in the partially filled $t_{2g}$-band. We will explicitly show how the lifting of the
orbital degeneracy by the monoclinic distortion stabilizes C-type antiferromagnetic (AFM) state,
which can be further transformed to the G-type AFM state by changing the crystal distortion from
monoclinic to orthorhombic one. Two microscopic mechanisms of such a stabilization,
associated with the one-electron crystal field and electron correlation interactions, are
discussed. The flexibility of the orbital degrees of freedom is analyzed in terms
of the magnetic-state dependence of interatomic magnetic interactions.
\end{abstract}

\section{Introduction}
\label{intro}

  The transition-metal oxides are currently regarded as some promising materials for
the future generation of electronic devises. The interest to these systems was spurred
by the discoveries in them of such key phenomena as
\begin{enumerate}
\item
high-temperature superconductivity (Cu- and Fe-based oxides)~\cite{Norman},
\item
colossal magnetoresistance (doped manganites)~\cite{Tokura},
\item
multiferroelectricity (BiMnO$_3$, TbMnO$_3$)~\cite{Khomskii},
\end{enumerate}
and many others, which can be potentially used in applications.
One important aspect of the transition-metal oxides is realization of the
so-called
``switching phenomena'', where electronic properties can be controlled
or easily switched between different states by various external factors, such as the
hydrostatic pressure, electric or magnetic field, etc. For instance, the
colossal magnetoresistance is the gigantic suppression of resistivity by
the magnetic field, the multiferroelectricity provides a unique possibility for
controlling the electric polarization by the magnetic field and vice versa, etc.

  Besides the applications, there exists a long-standing fundamental interest to the
transition-metal oxides, related to understanding of the
above-mentioned phenomena on the microscopic level.

  Historically,
theoretical developments around the transition-metal oxides
went almost parallel in two directions:
\begin{enumerate}
\item
model approaches, based on the solution and analysis of all possible
model Hamiltonians (the typical example is the Hubbard model, which is the basic and
widely used model in the physics of strongly correlated systems)~\cite{IFT};
\item
first-principles calculations, many of which are based on the
density-functional theory (DFT)~\cite{Kohn}.
\end{enumerate}
Each of these directions has merits and demerits.
For example, the main advantage of the model analysis is the simplicity
and transparency, while the main disadvantage is the use of adjustable parameters,
which are typically chosen to fit the experimental data. On the contrary,
the undisputable advantage of first-principles calculations is the
lack of adjustable parameters. However, as calculations become more and more
complex, we inevitably face the question about the interpretation of the
obtained results. The problem is complicated by the fact that many
first-principles methods are supplemented with some additional approximation
for the exchange-correlation interactions. The most typical one is the
local-density approximation (LDA), which is based on the model of
homogeneous electron gas and undermines the physics of
short-range Coulomb correlations. Thus, in the first-principles calculations
for the transition-metal oxides we frequently face the question: are the
obtained results physical or artifacts of additional approximations,
which were employed in the process of calculations?

  The primary goal of the new project, which is called ``realistic modeling'', is the construction of ``intelligent''
models, which would combine the principles of simplicity and transparency with the rigorous
first-principle basis for determination of the model parameters.

  What is the main idea of realistic modeling?

\begin{figure*}
  \includegraphics[width=0.75\textwidth]{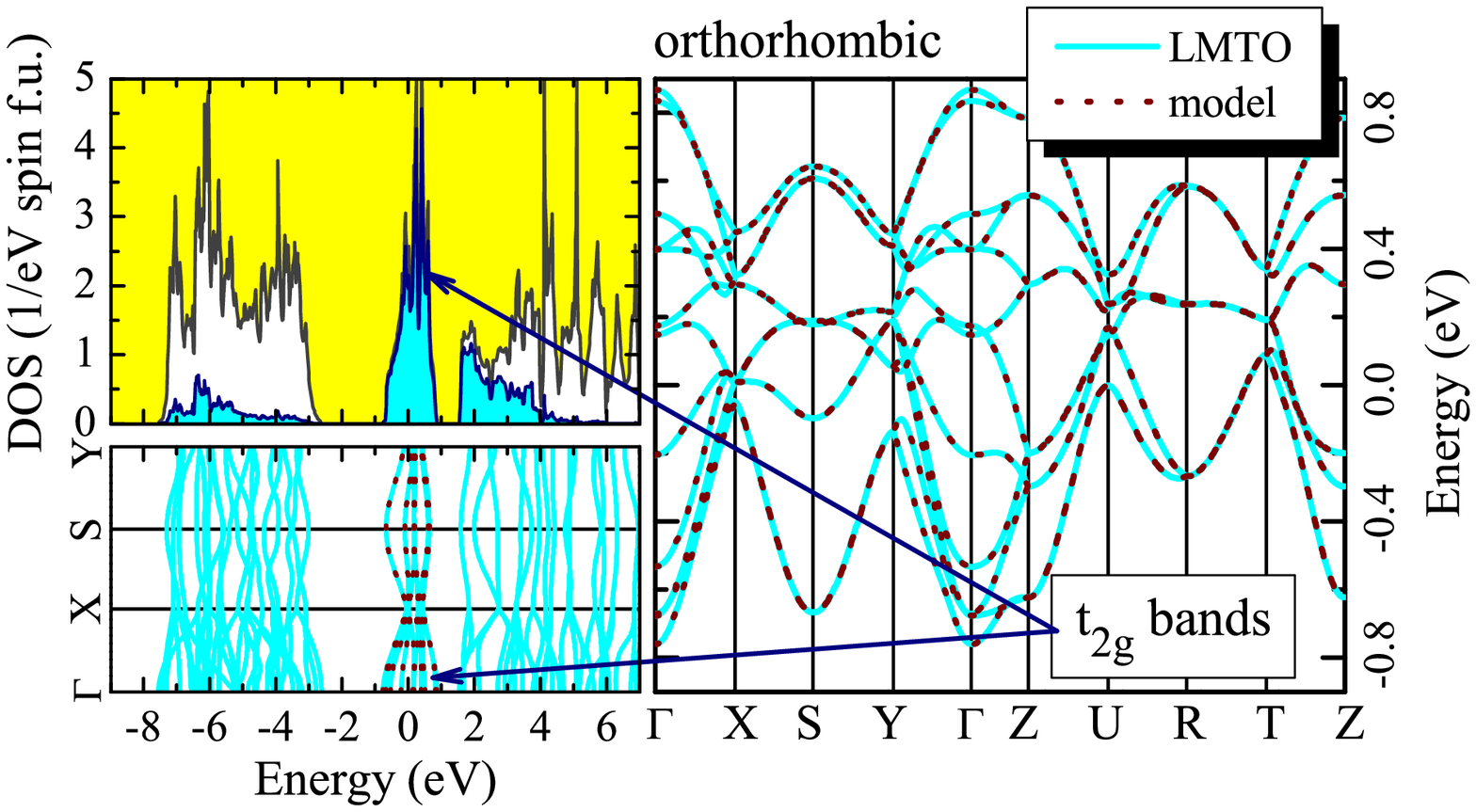}

  \includegraphics[width=0.75\textwidth]{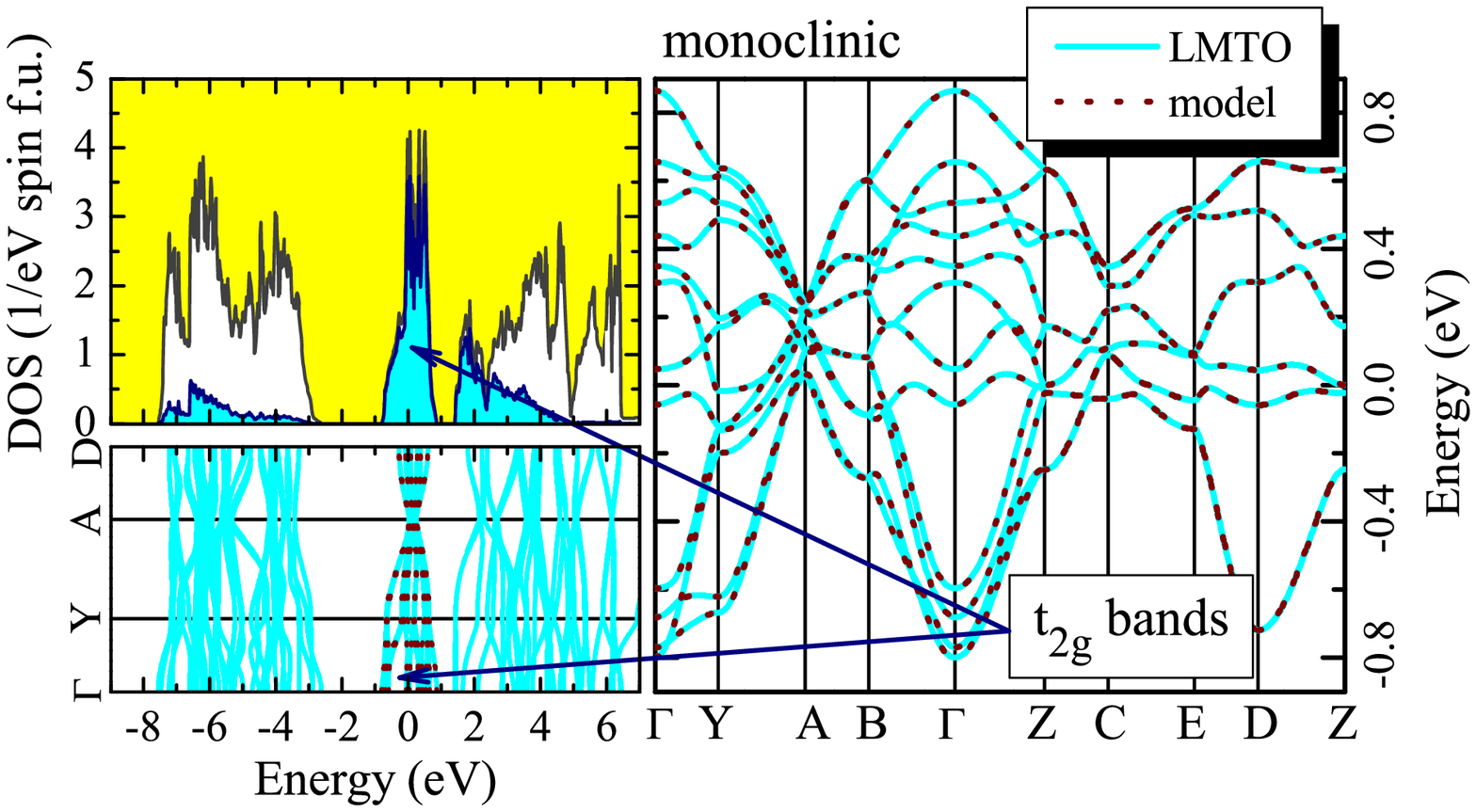}
\caption{Electronic structure for the
orthorhombic (top) and monoclinic (bottom) phases of SmVO$_3$ in the
local-density approximation.
The shaded area shows the contributions of the
V-$3d$ states. Right part of the figure shows
the enlarged behavior of the $t_{2g}$-bands
computed from the original LMTO basis functions~\protect\cite{LMTO} (solid curves)
and downfolded bands obtained in the process of construction of the low-energy model (dot-dashed curves).
The corresponding bands in the left part of the figure are shown by arrows.
The
Fermi level is at zero energy.}
\label{fig:LDA}       % Give a unique label
\end{figure*}

  Of course, the crystal and electronic structure of many oxide materials can be very complex.
Nevertheless, as long as we are mainly concerned with
electronic and magnetic properties of these systems, in many cases we can limit our consideration
to a small group of states
(the so-called low-energy states), which are
located near the Fermi level and which are mainly responsible for the considered properties.
The typical example of the LDA band structure of SmVO$_3$ is shown in Fig.~\ref{fig:LDA}.
In this case, the low-energy states are the $t_{2g}$-bands. Thus, the main idea is to
\begin{enumerate}
\item
construct an effective model for the low-energy states (typically, the multiorbital
Hubbard model);
\item
include the effect of all other states in some approximate form, through the
definition of the model parameters for the low-energy states;
\item
determine all model parameters rigorously, on the basis of first-principles
calculations of the electronic structure.
\end{enumerate}

  What are the main goals of the realistic modeling?
\begin{enumerate}
\item
It provides a possibility to go beyond the local-density approximation and to
systematically study the effects of electron-electron correlations in the narrow bands.
The solution of such a many-body problem is an extremely difficult task, even for the
present computational facilities. Therefore, if the is a chance to formulated
this problem rigorously
in some \textit{restricted} Hilbert space of states, near the Fermi level,
there is a chance to solve it, at least numerically.
\item
It is certainly true, that the first-principles calculations of the electronic
structure are currently on the rise. However, as the complexity of such calculations
also rises, we will inevitably need some tool for the analysis and interpretation
of the obtained results. We would like to emphasize that the final goal
of computational physics is not to reproduce some
complex
experimental trend. The final goal is to understand it and to come up with
some transparent explanation. From this point of view,
the realistic modeling will continue to play an important role, as a convenient
tool for the analysis and interpretation of results of complex calculations.
\end{enumerate}

  Finally, the construction of the low-energy model is
always conjugated with some approximations (and, as a matter of fact, the form of the
model itself is the main approximation). However, we would like to emphasize
from the very beginning
that
apart from these approximations, we do not use any adjustable parameters. Instead,
the realistic modeling brings the state of the discussion to a qualitatively new level:
if the model does not work, we need to reconsider the approximations
underlying
the definition of the model parameters or maybe the model itself.

\section{Construction of multiorbital Hubbard model for the low-energy states}
\label{model:construction}

  The first step of realistic modeling is the construction of an effective
multiorbital Hubbard model
in the low-energy part of spectrum:
\begin{equation}
\hat{\cal{H}}  =  \sum_{ij} \sum_{\alpha \beta}
t_{ij}^{\alpha \beta}\hat{c}^\dagger_{i\alpha}
\hat{c}^{\phantom{\dagger}}_{j\beta} +
  \frac{1}{2}
\sum_i  \sum_{\alpha \beta \gamma \delta} U_{\alpha \beta
\gamma \delta} \hat{c}^\dagger_{i\alpha} \hat{c}^\dagger_{i\gamma}
\hat{c}^{\phantom{\dagger}}_{i\beta}
\hat{c}^{\phantom{\dagger}}_{i\delta},
\label{eqn.ManyBodyH}
\end{equation}
which is regarded as the basic model in the physics of strongly correlated systems.
Other models can be derived by considering
some limiting cases of (\ref{eqn.ManyBodyH}) (an example of the Heisenberg model
will be considered in the next section).
The Hubbard model is specified
in the basis of Wannier orbitals, which are denoted by
Greek symbols, each of which is a
combination of spin
($s$$=$ $\uparrow$ or $\downarrow$) and orbital ($m$)
variables. The site-diagonal part of $\hat{t}_{ij}$
describes the crystal-field splitting of the atomic levels,
while
the off-diagonal ($i$$\ne$$j$) elements
stand for transfer integrals. If the relativistic spin-orbit interaction is not
included, $\hat{t}_{ij}$ is diagonal with respect
to the spin indices.
$U_{\alpha \beta \gamma \delta}$ are the matrix elements of screened
Coulomb interactions for the low-energy states.

  The most straightforward way to derive $\| \hat{t}_{ij} \|$ and
$\| U_{\alpha \beta \gamma \delta} \|$ is to use a complete basis of
localized orbitals in the low-energy part of spectrum. By the definition,
this is the basis of Wannier functions~\cite{Wannier}. There is a number of modern
techniques, which can be used for the construction of Wannier functions,
starting from one-electron Kohn-Sham orbitals in LDA.
In the basis of localized pseudoatomic orbitals, which are used for example in the linear muffin-tin orbital
(LMTO) method~\cite{LMTO}, it is convenient to use the projector-operator technique~\cite{PRB07}.
In the plane-wave basis, the Wannier functions can be constructed by minimizing the
matrix elements of some local operators,
such as the square of the position operator~\cite{MarzariVanderbilt}.

  Then, the one-electron part $\| \hat{t}_{ij} \|$ of the Hubbard model (\ref{eqn.ManyBodyH})
is identified with the matrix elements of the Kohn-Sham Hamiltonian in the basis of
Wannier functions.\footnote{Indeed, the Kohn-Sham Hamiltonian is also one-electron one.
Moreover, since the exchange correlation potential in LDA is local, it does not contribute
to the matrix elements between different atomic sites, so that the latter can be identified
with the kinetic transfer integrals. Meanwhile, the site-diagonal contributions of
the LDA potential to $\| \hat{t}_{ij} \|$ should be subtracted, because the splitting of the
atomic levels caused by electron-electron interactions is explicitly included in the
second part of the Hubbard model~\cite{review2008}.}
Since the Wannier basis is complete in the low-energy part of spectrum, the construction
is exact, and the one-electron energies derived from $\| \hat{t}_{ij} \|$ coincide with the
eigenenergies of the original LDA Hamiltonian (Fig.~\ref{fig:LDA}).
Historically, the derivation of
$\| \hat{t}_{ij} \|$ was based on the downfolding technique~\cite{PRB06a}.
However, after some modifications, aiming to get rid of the frequency-dependence
of $\hat{t}_{ij}$, this approach becomes equivalent to the
projector-operator technique~\cite{PRB07}.

  The effective Coulomb interaction in solids is defined as the energy cost for transferring
an electron from one atomic site to another:
\begin{equation}
2(d^n) \rightleftharpoons d^{n+1} + d^{n-1}.
\label{eqn.UBasic}
\end{equation}
Even if such redistribution of electrons occurs in the low-energy part of spectrum
(between appropriate Wannier orbitals),
the corresponding change of the electron density can interact with the other electrons
and
change the electronic structure in the entire range
of both low-energy and high-energy states. The change of the electronic structure in the high-energy
part contributes to the screening of Coulomb interactions $\| U_{\alpha \beta \gamma \delta} \|$
in the low-energy part. Some of these effects can be described in the
framework of the
random-phase approximation (RPA)~\cite{Ferdi04}, which takes into account the screening of
$U_{\alpha \beta \gamma \delta}$ due to the deformation of the Kohn-Sham orbitals in the
course of the reaction (\ref{eqn.UBasic}).

  Somewhat heuristically, in the pseudoatomic basis, the screening of Coulomb interactions,
associated
with the reaction (\ref{eqn.UBasic}),
can be divided in two parts \cite{PRB06a,review2008}:
\begin{enumerate}
\item
the screening, caused by relaxation of the pseudoatomic (basis) orbitals;
\item
the self-screening of the low-energy states by the atomic states of the same origin,
which contribute to other parts of spectrum due to the hybridization (the typical example is
the screening of the $t_{2g}$-bands of the $3d$-origin by the oxygen- and $e_g$-bands, which have a
large weigh of the atomic $3d$-states due to the hybridization effects -- see Fig.~\ref{fig:LDA}).
\end{enumerate}
These two channel of screening can be easily incorporated in the framework of the constrained
density-functional theory (DFT) and RPA, respectively~\cite{PRB06a}. Such a separation considerably
facilitates the calculations and make them more transparent.

\section{Solution of the model Hamiltonian}
\label{model:solution}
The simplest way to solve the many-body problem (\ref{eqn.ManyBodyH}) is to use the
mean-field Hartree-Fock (HF) approach, where the ground-state wavefunction
is approximated by the single Slater determinant, constructed from the one-electron
orbitals $\{ \phi_k \}$. The latter are obtained from the solution of the one-electron
equations (in the reciprocal space):
\begin{equation}
\left( \hat{t}_{\bf k} + \hat{\cal V} \right) | \varphi_{k} \rangle =
\varepsilon_k | \varphi_k \rangle,
\label{eqn:HFeq}
\end{equation}
where $\hat{t}_{\bf k}$ is the Fourier image of $\hat{t}_{ij}$,
$k$ is a collective
index combining the momentum
${\bf k}$ of the first Brillouin zone, the band number, and
the spin ($s$$=$ $\uparrow$ or $\downarrow$) of the particle, and
$\hat{\cal V}$ is the HF potential
\begin{equation}
{\cal V}_{\alpha \beta} = \sum_{\gamma \delta}
\left( U_{\alpha \beta \gamma \delta} - U_{\alpha \delta \gamma \beta} \right) n_{\gamma \delta},
\label{eqn:HFpot}
\end{equation}
expressed through the density matrix
$$
\hat{n} \equiv \| n_{\gamma \delta} \| = \sum_k^{occ} | \varphi_k \rangle \langle \varphi_k |.
$$
The latter is obtained self-consistently. By knowing $\{ \phi_k \}$ and $\{ \varepsilon_k \}$,
it is easy to construct the one-electron (retarded) Green function
$\hat{\cal G}^{\uparrow, \downarrow}_{ij}(\omega)$ for the spin $\uparrow$ and $\downarrow$,
which can be used in many applications. Particularly, one useful application is
related to
the analysis of interatomic magnetic interactions, which describe infinitesimal rotations
of the spins near different magnetic equilibriums~\cite{Liechtenstein}.
In this case, the total energy change can be
mapped onto the Heisenberg model
\begin{equation}
E_{\rm Heis} = -\frac{1}{2} \sum_{ij} J_{ij} {\bf e}_i \cdot {\bf e}_j
\label{eqn:HHeisenberg}
\end{equation}
(${\bf e}_i$ and ${\bf e}_j$ being the \textit{directions} of the spin moments),
and the parameters are given by
\begin{equation}
J_{ij} = \frac{1}{2 \pi} {\rm Im} \int_{-\infty}^{\varepsilon_{\rm F}}
d \omega {\rm Tr}_L \left\{ \hat{\cal G}_{ij}^\uparrow (\omega)
\Delta \hat{\cal V} \hat{\cal G}_{ji}^\downarrow (\omega)
\Delta \hat{\cal V} \right\}.
\label{eqn:JHeisenberg}
\end{equation}
In these notations,
$\Delta \hat{\cal V} = {\rm Tr}_S \{ \hat{\sigma}_z \hat{\cal V} \}$
is the spin part of the Hartree-Fock potential,
${\rm Tr}_S$ (${\rm Tr}_L$) denotes the trace over the spin (orbital) indices,
and
$\hat{\sigma}_z$ is the Pauli matrix.\footnote{More precisely,
the parameters
$J_{ij}$ in (\ref{eqn:JHeisenberg}) define the \textit{local} stability of the
given
magnetic configuration, which is stable if $J_{ij}>0$ and unstable if $J_{ij}<0$.
In order to be consistent with the \textit{global}
definition of the parameters of the Heisenberg model (\ref{eqn:HHeisenberg}), they should be
additionally multiplied by ${\bf e}_i \cdot {\bf e}_j$$=$ $1$ and $-$$1$, correspondingly
for the ferromagnetic and antiferromagnetic bonds.
}

  The simplest way to go beyond the HF approximation is to consider the regular perturbation theory
for correlation interactions. The latter are defined as the difference between true
operator of
electron-electron interactions in the Hubbard model (\ref{eqn.ManyBodyH}) and its
counterpart in the HF approximation (i.e., the combination of Coulomb and exchange
interactions):
\begin{equation}
\hat{\cal{H}}_C = \sum_i \left(
\frac{1}{2} \sum_{\alpha \beta \gamma \delta}
U_{\alpha \beta \gamma \delta}
\hat{c}^\dagger_{i\alpha} \hat{c}^\dagger_{i\gamma}
\hat{c}^{\phantom{\dagger}}_{i\beta} \hat{c}^{\phantom{\dagger}}_{i\delta} -
\sum_{\alpha \beta} {\cal V}_{\alpha \beta}
\hat{c}^\dagger_{i\alpha} \hat{c}^{\phantom{\dagger}}_{i\beta} \right).
\label{eqn:Hcorrelations}
\end{equation}
For instance, in the second order, the correlation energy is given by:
\begin{equation}
E_C^{(2)} = - \sum_{S} \frac{
\langle G | \hat{\cal{H}}_C | S \rangle \langle S | \hat{\cal{H}}_C | G \rangle }
{E_{\rm HF}(S) - E_{\rm HF}(G)},
\label{eqn:dE2ndorder}
\end{equation}
where $|G \rangle$  is the ground-state wavefunction in the HF approximation and
$|S \rangle$ is the excited state, which is obtained from $|G \rangle$ by replacing two
orbitals in the occupied part of spectrum by two unoccupied orbitals.

Of course, this strategy is not universe. Nevertheless, it can justified when the
degeneracy of the ground state is already lifted (for instance, by the lattice distortions),
so that the ground state can be described reasonably well by the single Slater determinant
in the HF approximation and the correlation interactions can be included after that by
considering the regular (non-degenerate) perturbation theory expansion.

In principle, one can go beyond the second order and consider higher-order effects
for correlations interactions in the framework of the T-matrix theory \cite{Kanamori,JETP07}.

\section{Applications to vanadates $R$VO$_3$ ($R$$=$ La, Ce, Pr, Nd, Sm, Gd, Tb, Yb, and Y)}
\label{applications}

\begin{figure*}
  \includegraphics[width=0.75\textwidth]{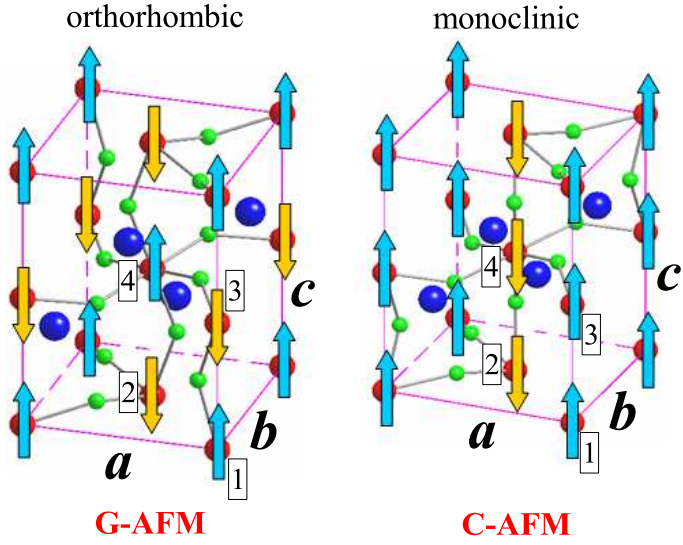}
\caption{G- and C-type antiferromagnetic ordering realized in the orthorhombic
and monoclinic structure of YVO$_3$.}
\label{fig:MagneticStructure}       % Give a unique label
\end{figure*}

  The vanadates $R$VO$_3$ have attracted a considerable experimental and theoretical
attention. All these compounds crystallize in the highly distorted perovskite structure.
However, even small change of the crystal structure in $R$VO$_3$ can lead to
dramatic change of the magnetic structure. Thus, this behavior can be regarded as a
prototype of the ``switching phenomena''. Below the magnetic transition temperature,
the crystal structure of $R$VO$_3$ can be
either orthorhombic (the space group $D_{2h}^{16}$) or monoclinic
(the space group $C_{2h}^5$). The orthorhombic structure typically coexists with the
G-type antiferromagnetic (AFM) ordering (Fig.~\ref{fig:MagneticStructure}),
where all nearest-neighbor V-sites are coupled
antiferromagnetically, while the monoclinic structure coexists with the C-type AFM
ordering (ferromagnetic chains propagating along the ${\bf c}$-axis and
antiferromagnetically coupled in the ${\bf ab}$-plane).

  Depending on the size of the $R^{3+}$ ions, which control the magnitude of the
lattice distortion (particularly, the angles V-O-V, etc.), one can distinguish
three types of behavior of $R$VO$_3$:
\begin{enumerate}
\item
The materials containing large ions, such as La$^{3+}$, Ce$^{3+}$, Pr$^{3+}$, and Nd$^{3+}$,
below the magnetic transition point crystallize in the monoclinic structure and
develop the C-type AFM ordering. The transition to the orthorhombic phase typically occurs
right above the
N\'{e}el temperature ($T_{\rm N}$)~\cite{Zubkov,Bordet};
\item
The low-temperature properties of $R$VO$_3$ with $R$$=$ Sm, Gd, and Tb
are marked by a coexistence of orthorhombic and monoclinic
phases. The monoclinic phase is formed above $T_{\rm N}$.
The orthorhombic phase starts to develop near
magnetic transition and its proportion gradually increases with
decreasing temperature~\cite{SagePRL,SageThesis,Yusupov}. For example, at $T$$=$ 5 K,
the proportion of the orthorhombic fraction is 25~\% in SmVO$_3$~\cite{SagePRL} and
about 70~\% in GdVO$_3$~\cite{Yusupov}.
\item
In YVO$_3$, the monoclinic structure is also formed well above $T_{\rm N}$$=$ 116 K.
With decreasing temperature, first C-type AFM ordering develops below $T_{\rm N}$,
within given monoclinic symmetry.
Then, at $T_s$$=$ 77 K the crystal structure suddenly changes
to the orthorhombic one.
The corresponding magnetic structure also changes from C- to G-type AFM~\cite{Blake}.
Similar behavior is observed in YbVO$_3$.
\end{enumerate}

  Details of the experimental crystal structure and the magnetic properties
of $R$VO$_3$ can be found in \cite{Zubkov,Bordet,SagePRL,SageThesis,Blake}. In the present work, we will use
distorted vanadates
in order
to illustrate the main ideas of realistic modeling. There are several reasons why
we have selected $R$VO$_3$ for these purposes:
\begin{enumerate}
\item
The microscopic origin of two magnetic structures, which are realized in $R$VO$_3$,
is still under debates. Although main details of the magnetic structures can be
ascribed to the lattice distortions, which restrict the variation of
the orbital degrees of freedom near
some particular configurations~\cite{SawadaTerakura,FangNagaosa}, the role of
orbital fluctuations away from these configurations is also actively discussed
in the literature~\cite{Khaliullin}. Particularly, the magnetic interactions in these systems
are organized in such a way that the spin degrees of freedom are tightly coupled to
the orbital ones, where any change of the orbital configuration affects the
magnetic structure and vice versa. In the present work, we will illustrate
how the computational tools, introduced in Sec.~\ref{model:solution},
can be used for the analysis of these properties.
\item
The vanadates $R$VO$_3$ are currently regarded as some test compounds for
various theories aiming to describe the close interplay among spin, orbital, and
lattice degrees of freedom. Therefore, it is very important to test the ideas
of realistic modeling on these systems and to apply them to the whole series
of compounds $R$VO$_3$
with different $R$. All previous studies were mainly limited by
LaVO$_3$ and YVO$_3$~\cite{review2008,SawadaTerakura,FangNagaosa,PRB06b}.
\item
We will also use $R$VO$_3$ in order to discuss some prototypical examples of the
``switching phenomena'', which could be important in applications. Particularly,
we will show how the magnetic structure of $R$VO$_3$ can be switched by
changing the
lattice distortions.
An interesting example of switching between different spin and orbital states in $R$VO$_3$
by optical pulses was considered recently in \cite{Yusupov,Tomimoto,Mazurenko}.
Another interesting example is the ``magnetic field switching'', realized in
DyVO$_3$~\cite{Miyasaka}.
\end{enumerate}

\subsection{Parameters of the model Hamiltonian}
\label{ParametersOfModel}

  In all the applications, the model was constructed for the $t_{2g}$-bands,
located near the Fermi level (see Fig.~\ref{fig:LDA}).\footnote{
We did not consider the magnetic effects associated with the
rare-earth $4f$-states, which can be also interesting and important~\cite{Reehuis}.
It could be a good subject for a separate study, but in the present work,
the $4f$-states
were treated as non-polarized quasiatomic core states.
}
In vanadates $R$VO$_3$,
each of V-site donates two electrons to these bands.

  The first important set of parameters is related to the crystal field (CF), which is
specified by the site-diagonal elements $\hat{t}_{ii}$ of the one-electron Hamiltonian.
It breaks the degeneracy of the atomic levels and define
the type of occupied $t_{2g}$-orbitals in the atomic limit.
The interactions, which may act against the crystal field and deform the
occupied orbitals are controlled by the intraatomic
Coulomb repulsion ${\cal U}$, Hund's rule
coupling ${\cal J}$ and transfer integrals $\hat{t}_{ij}$ between
different atomic sites. They will be discussed below.

\begin{figure*}
  \includegraphics[width=0.75\textwidth]{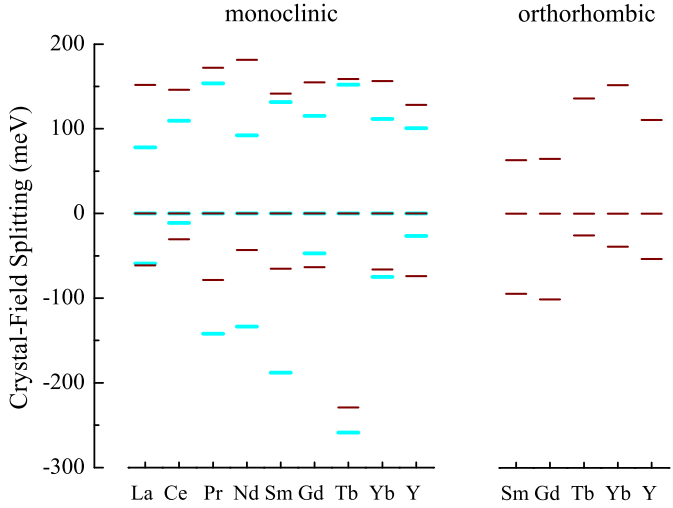}
\caption{The crystal-field splitting of three $t_{2g}$-levels
in
monoclinic (left) and
orthorhombic (right) phases of $R$VO$_3$. The energies are measured
relative to the middle level, corresponding to the highest occupied orbital
in the atomic limit. The splitting for two vanadium sublattices
in the monoclinic phase
is shown by different colors, where parameters for the less distorted sublattice
(according to the energy difference between middle and highest orbitals) are
denoted by the light blue color, while the ones for the more distorted sublattice --
by the dark brown color.}
\label{fig:CF}       % Give a unique label
\end{figure*}

  The scheme of the $t_{2g}$-level splitting, obtained from the
diagonalization of $\hat{t}_{ii}$, is shown in Fig.~\ref{fig:CF}.
The physically relevant parameter is the splitting between middle
and highest orbitals
(corresponding to the highest occupied and lowest
unoccupied orbitals in the atomic limit). In the most cases this splitting is
at least 100 meV. Somewhat exceptional behavior is observed in the monoclinic phase of LaVO$_3$
and in the orthorhombic phases of SmVO$_3$ and GdVO$_3$, were the characteristic
splitting can be as small as 60-80 meV. Some consequences of this
behavior will be discussed below. In the monoclinic phase,
the splitting can be very different in two inequivalent V-sublattices,
which are denoted as (1,2) and (3,4) in Fig.~\ref{fig:MagneticStructure}.
For example, in LaVO$_3$ this splitting is 78 meV and 152 meV, respectively.
An example of the CF-orbitals, obtained after the diagonalization of $\hat{t}_{ii}$
for SmVO$_3$,
is presented in Table~\ref{tab:CF}.
\begin{table}
\caption{Eigenenergies (measured in meV from the middle level)
and eigenvectors obtained from the diagonalization
of the crystal-field Hamiltonian $\hat{t}_{ii}$
for orthorhombic (`o') and monoclinic (`m') phases
of SmVO$_3$.
The eigenvectors are expanded over the basis of
$xy$, $yz$, $z^2$, $zx$, and $x^2$-$y^2$ Wannier-orbitals,
in the orthorhombic coordinate frame.
Positions of atomic sites are explained in Fig.~\protect\ref{fig:MagneticStructure}.
The eigenvectors for other sites can be obtained using symmetry operations.}
\label{tab:CF}       % Give a unique label
\begin{tabular}{ccrc}
\hline\noalign{\smallskip}
phase & site & energies & orbitals  \\
\noalign{\smallskip}\hline\noalign{\smallskip}
 & & $-95$ \ \ \ &
$\phantom{-0.0}0, \phantom{-}0.22, -0.39, \phantom{-}0.84, \phantom{-}0.30$
\\
`o'  & $1$ & $0$ \ \ \ &
$\phantom{-}0.06, \phantom{-}0.78, \phantom{-}0.22, -0.28, \phantom{-}0.51$
\\
 & & $63$ \ \ \ &
$\phantom{-}0.32, \phantom{-}0.52, \phantom{-}0.06, \phantom{-}0.17, -0.77$
\\
\noalign{\smallskip}
 & & $-65$ \ \ \ &
$-0.07, \phantom{-}0.70, \phantom{-}0.30, \phantom{-}0.63, -0.14$
\\
`m'  & $1$ & $0$ \ \ \ &
$-0.37, -0.05, \phantom{-}0.03, \phantom{-}0.20, \phantom{-}0.91$
\\
 & & $142$ \ \ \ &
$\phantom{-}0.12, \phantom{-}0.62, \phantom{-}0.18, -0.72, \phantom{-}0.23$
\\
\noalign{\smallskip}
 & & $-188$ \ \ \ &
$-0.39, \phantom{-}0.02, \phantom{-}0.04, \phantom{-}0.05, \phantom{-}0.92$
\\
`m'  & $3$ & $0$ \ \ \ &
$\phantom{-}0.12, -0.87, \phantom{-}0.30, \phantom{-}0.37, \phantom{-}0.04$
\\
 & & $132$ \ \ \ &
$\phantom{-}0.20, \phantom{-}0.34, -0.21, \phantom{-}0.89, \phantom{-}0.04$
\\
\noalign{\smallskip}\hline
\end{tabular}
\end{table}

  An example of the transfer integrals in the local coordinate frame,
corresponding to the diagonal representation of the crystal field at each
atomic site, is shown in Table~\ref{tab:transferCF} for SmVO$_3$.
\begin{table}
\caption{Transfer integrals (measured in meV) between nearest neighbors in the
orthorhombic (`o') and monoclinic (`m') phases of SmVO$_3$
in the local coordinate frame corresponding to the diagonal
representation of the crystal field.
Positions of atomic sites are explained in Fig.~\protect\ref{fig:MagneticStructure}.}
\label{tab:transferCF}       % Give a unique label
\begin{tabular}{cccc}
\hline\noalign{\smallskip}
phase & $\hat{t}_{12}$ & $\hat{t}_{34}$ & $\hat{t}_{13}$ \\
\noalign{\smallskip}\hline\noalign{\smallskip}
`o' &
$
\left(
\begin{array}{rrr}
  69 &  157 &    4 \\
  2  & -86  & -94  \\
  25 &  25  & -123 \\
\end{array}
\right)
$
&
$
\left(
\begin{array}{rrr}
  69 &  157 &    4 \\
  2  & -86  & -94  \\
  25 &  25  & -123 \\
\end{array}
\right)
$
&
$
\left(
\begin{array}{rrr}
  86 & -53 &  -4 \\
 -53 &  56 & -64 \\
  -4 & -64 &  12 \\
\end{array}
\right)
$
\\
\noalign{\smallskip}
`m' &
$
\left(
\begin{array}{rrr}
  42 & -29  & -142 \\
 -36 & -118 & -40  \\
 -5  &  23  & -43  \\
\end{array}
\right)
$
&
$
\left(
\begin{array}{rrr}
 -121 & -43  & -24 \\
  41  &  22  & -12 \\
 -19  & -124 & -10 \\
\end{array}
\right)
$
&
$
\left(
\begin{array}{rrr}
  54 &  23  & -155 \\
  15 & -25  &   74 \\
 -21 & -139 & -102 \\
\end{array}
\right)
$
\\
\noalign{\smallskip}\hline
\end{tabular}
\end{table}
The form of these integrals is very different from the one
expected for the undistorted cubic structure.\footnote{
In the cubic structure, the hoppings are allowed only between alike orbitals,
lying
in the plane of the bond. For example, in the $z$-direction,
the hoppings can take place only between
$t_{2g}$-orbitals of either $yz$ or $zx$ symmetry~\cite{SlaterKoster}.
}
Thus, the crystal distortions have a profound effect not only on the
$t_{2g}$-level splitting, but also on the
form of the transfer integrals. All these details are
very
important for the analysis of the low-energy properties of vanadates.
Without help of the first-principles electronic structure calculations,
it is practically impossible to fix all the parameters in an
unambiguous manner.

\begin{figure*}
  \includegraphics[width=0.75\textwidth]{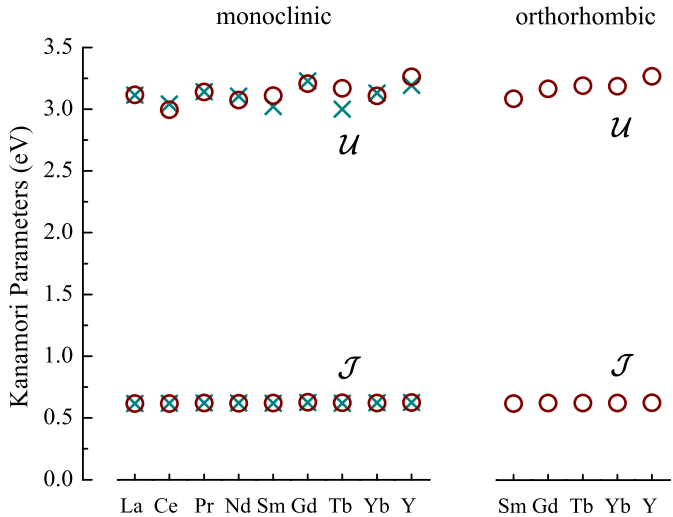}
\caption{Parameters of screened intraorbital Coulomb interaction ${\cal U}$
and the exchange interaction ${\cal J}$ derived for the $t_{2g}$-bands of $R$VO$_3$
in
monoclinic (left) and
orthorhombic (right) phase. The parameters for two
different V-sublattices in the monoclinic phase
are shown by different symbols: the values corresponding to the more distorted
environment (according to the scheme of the crystal-field splitting in Fig.~\protect\ref{fig:CF}) are
shown by circles, while the values corresponding to the less distorted environment
are shown by crosses.}
\label{fig:UJ}       % Give a unique label
\end{figure*}

  The behavior of screened Coulomb and exchange interactions in the $t_{2g}$-band
is illustrated in Fig.~\ref{fig:UJ}.
For these (purely explanatory) purposes, the whole matrices of
screened Coulomb interactions were fitted in terms of two Kanamori parameters~\cite{Kanamori}:
the intraorbital Coulomb interaction ${\cal U}$ and the exchange interaction ${\cal J}$.\footnote{
For third Kanamori parameter -- the interorbital Coulomb interaction ${\cal U}'$ --
we used the relation ${\cal U}' = {\cal U} - 2{\cal J}$, which holds for the $t_{2g}$-states in the cubic environment.
}
The Coulomb interaction ${\cal U}$ is slightly larger than 3 eV and somewhat sensitive to
the local environment of the V-sites in solids.
In orthorhombic systems, ${\cal U}$ increases with
the lattice distortion in the direction Sm$\rightarrow$Y. In monoclinic systems, this dependence
is not so obvious. Also in the monoclinic systems, the values of ${\cal U}$ are slightly different
for the inequivalent V-sublattices.
On the other hand, the screening of the exchange interaction ${\cal J}$ is practically insensitive to the
local environment, and for all considered compounds ${\cal J}$ is close to
$0.62$ eV.

  Another important parameter, which controls the orbital state and can compete
with the CF splitting, is the energy gain caused by the superexchange (SE) processes:
${\rm Tr}_L ( \hat{t}_{ij}\hat{t}_{ji} )/({\cal U}-3{\cal J})$ and
${\rm Tr}_L ( \hat{t}_{ij}\hat{t}_{ji} )/{\cal U}$, correspondingly for the
ferromagnetically and antiferromagnetically coupled bonds~\cite{KugelKhomskii}.
Using the values of transfer integrals between
nearest neighbors (Table~\ref{tab:transferCF}), Coulomb ${\cal U}$
and exchange ${\cal J}$ interactions (Fig.~\ref{fig:UJ}), this energy gain
can be estimated (very roughly) as 7-40 meV per one V-V bond.
Thus, the energy gain caused by the SE processes is expected to be smaller
or at least comparable with the CF splitting (Fig.~\ref{fig:CF}).\footnote{
Note that since each V-site participate in six bonds, the energy gain recalculated
``per site'' will be larger. On the other hand, the simple estimates in this
section do not take into account the Pauli principle, which forbids the hoppings
onto already occupied orbitals and makes energy gain smaller. Thus, this is
indeed only an ``order or magnitude estimate''.
}
Thus, there is no predominant mechanism and both crystal field
and SE processes can play some role in the formation of the
orbital (and related to it magnetic) ground state of vanadates $R$VO$_3$.
More detailed analysis will be given in the next sections.

\subsection{Relative stability of magnetic structures: total energies}
\label{TotalEnergies}

  In this section we will discuss abilities of theories,
which implies that the orbital degrees of freedom are frozen in some peculiar configuration by the lattice distortions,
and
the magnetic structure simply follow them,
basically according to the conventional Goodenough-Kanamori~\cite{GoodenoughBook,Kanamori1959}.\footnote{
In the present contents, the ``Goodenough-Kanamori rules'' means that the maximal overlap
between occupied orbitals in some particular bond favors AFM interactions, while
the minimal overlap between occupied orbitals (or
the maximal overlap between occupied and unoccupied orbitals) favors FM interactions.
}

\begin{figure*}
  \includegraphics[width=0.75\textwidth]{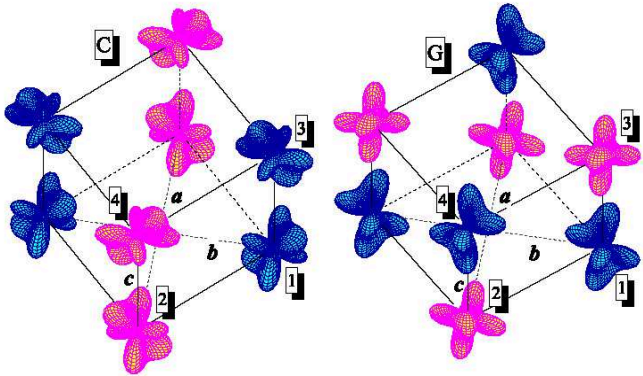}
\caption{Distribution of charge densities associated with occupied
$t_{2g}$-orbitals (the orbital ordering) realized in the monoclinic (left)
and orthorhombic (right) phases of TbVO$_3$ in the Hartree-Fock approximation.
Different magnetic sublattices associated with opposite
directions of spins in the C- (left) and G- (right) type antiferromagnetic
ground state are shown by different colors.}
\label{fig:OrbitalOrderTb}       % Give a unique label
\end{figure*}

  A typical example of the orbital ordering in shown in Fig.~\ref{fig:OrbitalOrderTb} for TbVO$_3$.
We have selected this material because it has the largest CF splitting (Fig.~\ref{fig:CF})
and therefore it is the most appropriate for the analysis in the present contents.
In the orthorhombic phase of TbVO$_3$, the occupied orbitals have a large overlap in all three
directions. Therefore, it is reasonable to expect the AFM character of interactions
and the G-type AFM ground state. In the monoclinic phase, the overlap between occupied
orbitals along the ${\bf c}$-axis is considerably weaker. Thus, this orbital
configuration will favor FM character of interactions in the ${\bf c}$-direction
and the C-type AFM ground state.

  Another important point of the CF theory is that as long as the orbitals degeneracy
is lifted by the lattice distortions, there is a hope that
the correct magnetic ground state can be reproduced
already at the level of the HF approximation. In some sense, the situation
is similar to the closed atomic shell, where the ground state is also non-degenerate
and can be described by the single Slater determinant. Then, if necessary, the correlation
interactions can be taken into account on the top of the HF approximation, using
regular perturbation theory expansion.

\begin{figure*}
  \includegraphics[width=0.375\textwidth]{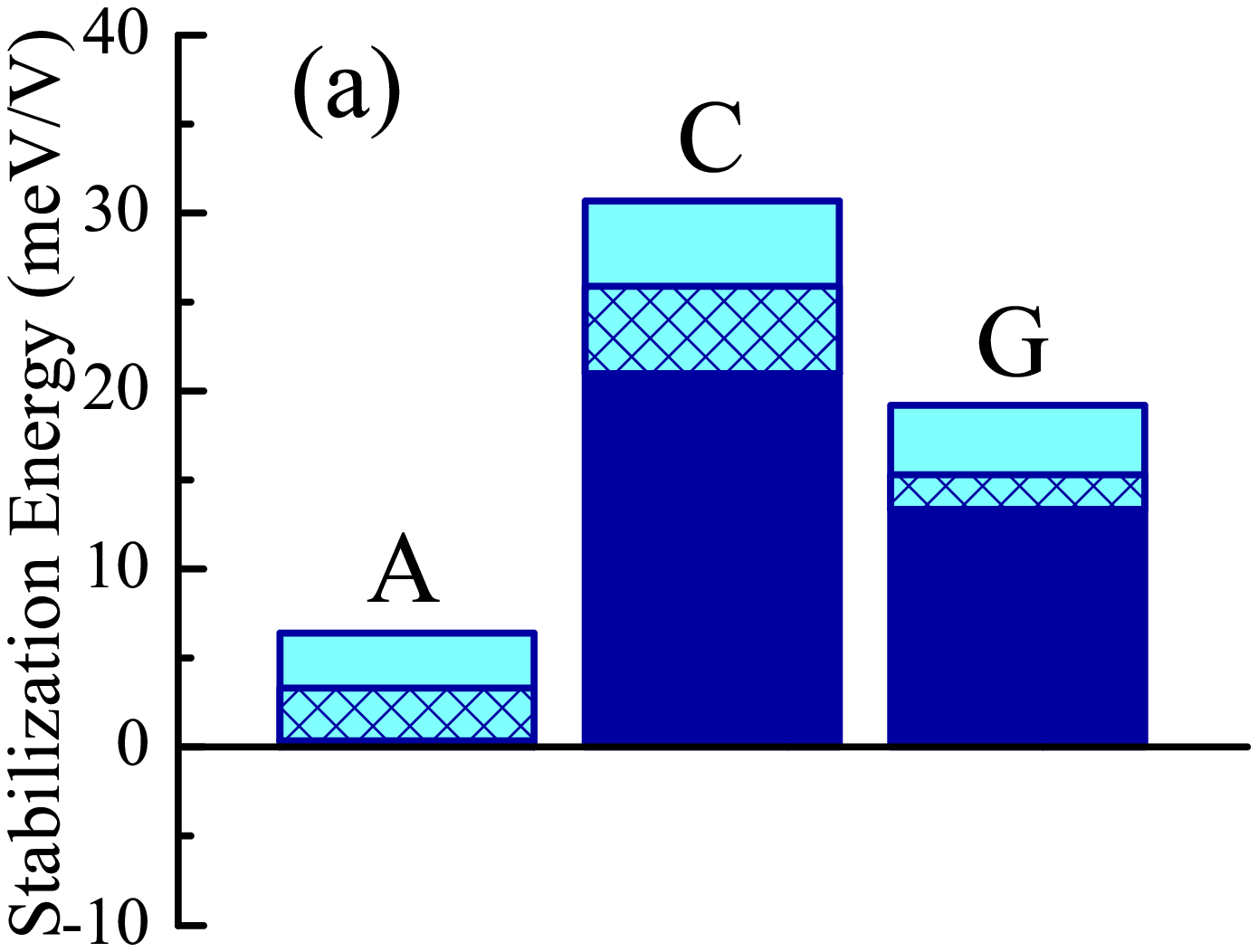}
  \includegraphics[width=0.375\textwidth]{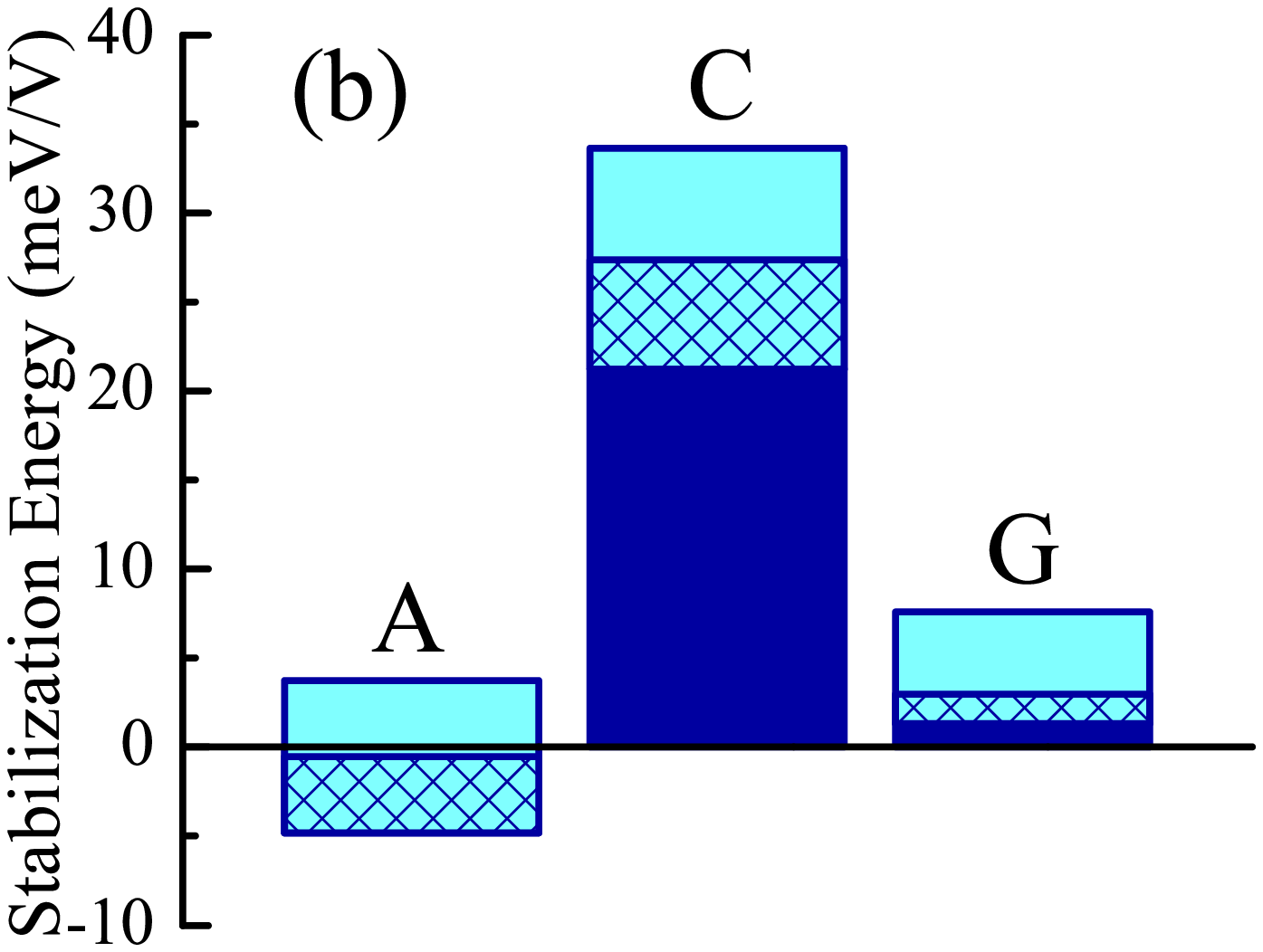}

  \includegraphics[width=0.375\textwidth]{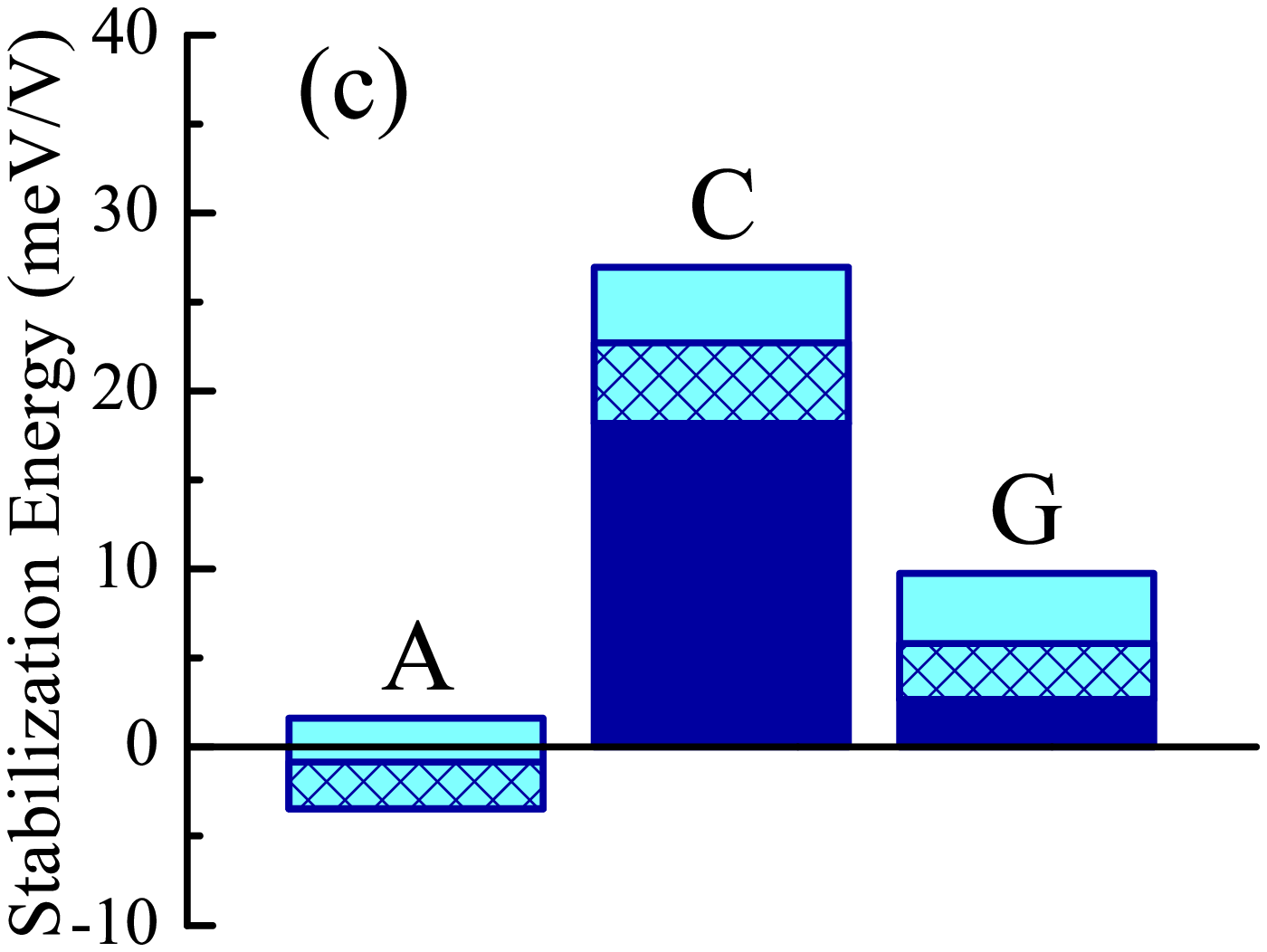}
  \includegraphics[width=0.375\textwidth]{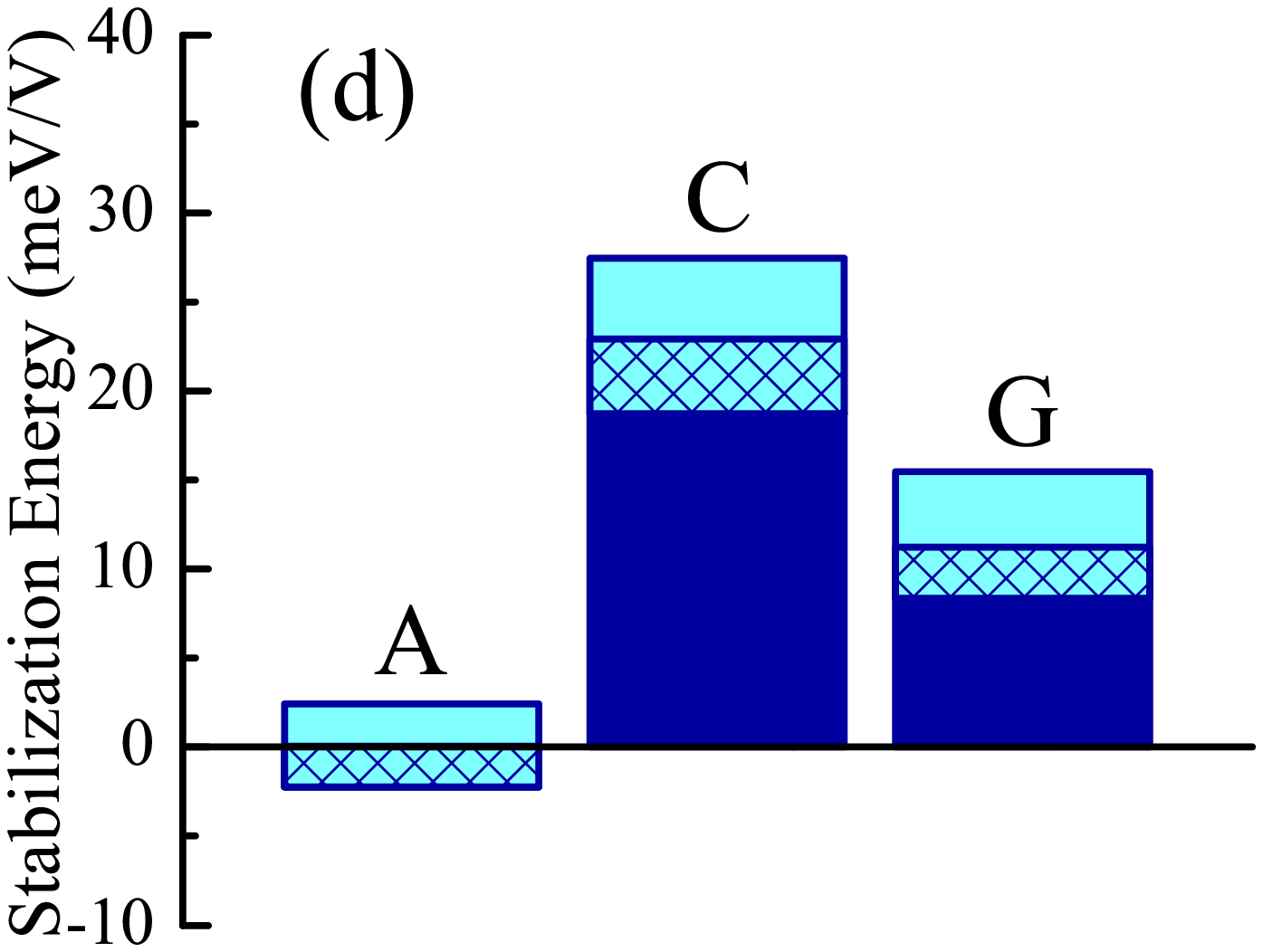}
\caption{Stabilization energies of the main antiferromagnetic states in
LaVO$_3$ (a), CeVO$_3$ (b), PrVO$_3$ (c), and NdVO$_3$ (d)
measured
relative to the ferromagnetic state as obtained in the Hartree-Fock approximation (dark blue area)
and after taking into account the correlation interactions in the second order of
perturbation theory (light blue area) and in the
framework of the T-matrix theory (hatched area).}
\label{fig:TotalEnergyLaCePrNd}       % Give a unique label
\end{figure*}

  All these trends are clearly seen in the total energy calculations for the $R$VO$_3$ ($R$$=$ La-Nd)
compounds, which below the magnetic transition temperature stabilize
in only one -- monoclinic structure. Fig.~\ref{fig:TotalEnergyLaCePrNd} shows the behavior
of stabilization energy (minus total energy measured relative to the FM state) for
the AFM configurations of the A-, C-, and G-type.\footnote{
The C- and G-type AFM structures are explained in Fig.~\ref{fig:MagneticStructure}.
The A-type AFM structure consists of the FM $\bf{ab}$-planes, which are coupled
antiferromagnetically along the ${\bf c}$-axis.
}
We note the following:
\begin{enumerate}
\item
The HF approximation predicts the correct
C-type AFM ground state for all four compounds;
\item
The correlation interactions, which were taken into account using regular perturbation
theory expansion near the HF solution for each magnetic state, additionally stabilize the experimentally observed
C-type AFM ordering.
Moreover, both second order perturbation and the T-matrix theory
provide very consistent explanation for the behavior of correlation energies,
although the
energies obtained in the T-matrix theory for the AFM states are
systematically smaller due to the higher-order correlation effects, which are
included to the T-matrix, but not to the second-order perturbation theory.
\end{enumerate}
Thus, not only the one-electron crystal field, which break the degeneracy
of the HF states in some peculiar way,
but also the correlation interactions play an important role in stabilizing
the experimentally observed C-type AFM ordering in compounds
$R$VO$_3$ ($R$$=$ La-Nd). The stabilization energies of
the
correlation origin are systematically larger for the C-type AFM state, irrespectively
on the approximation employed for treating the correlation interactions.

  How general are these results?

\begin{figure*}
  \includegraphics[width=0.75\textwidth]{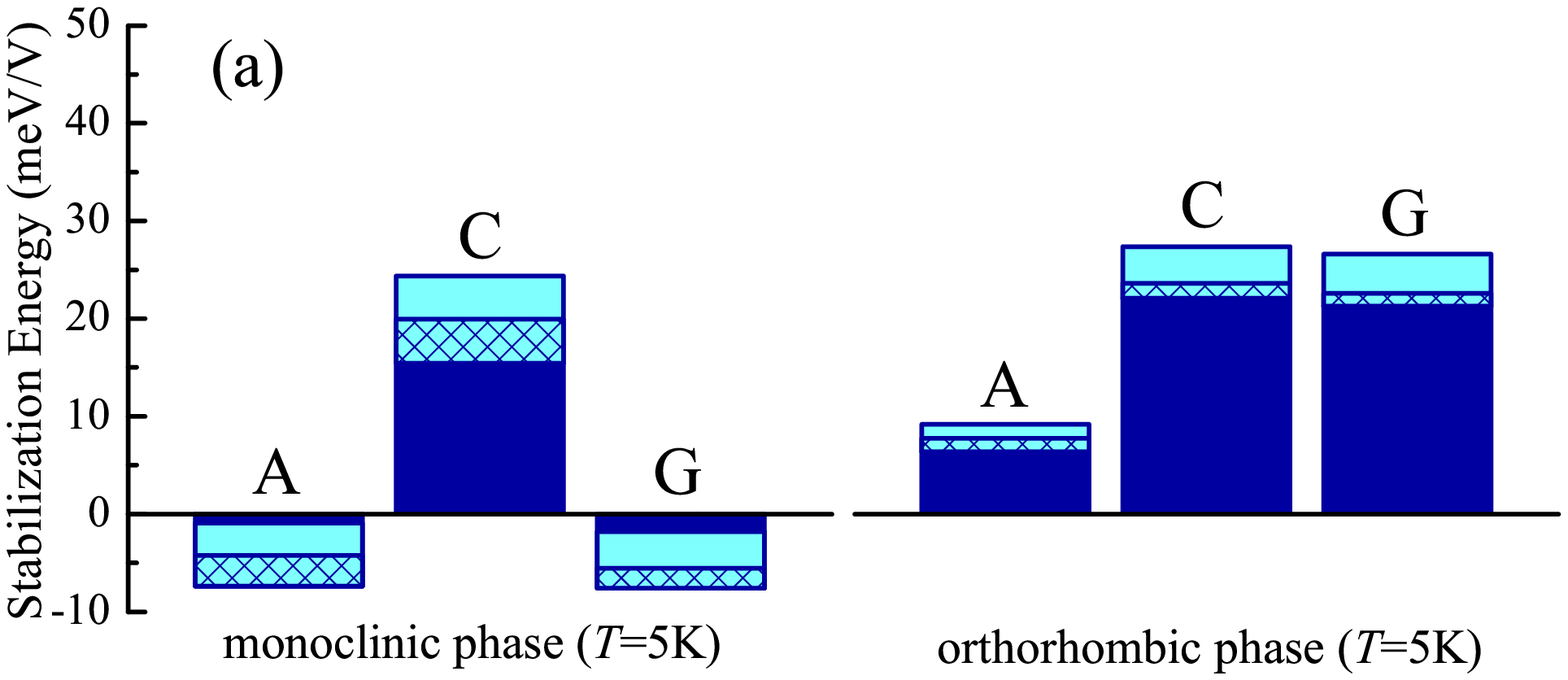}

  \includegraphics[width=0.75\textwidth]{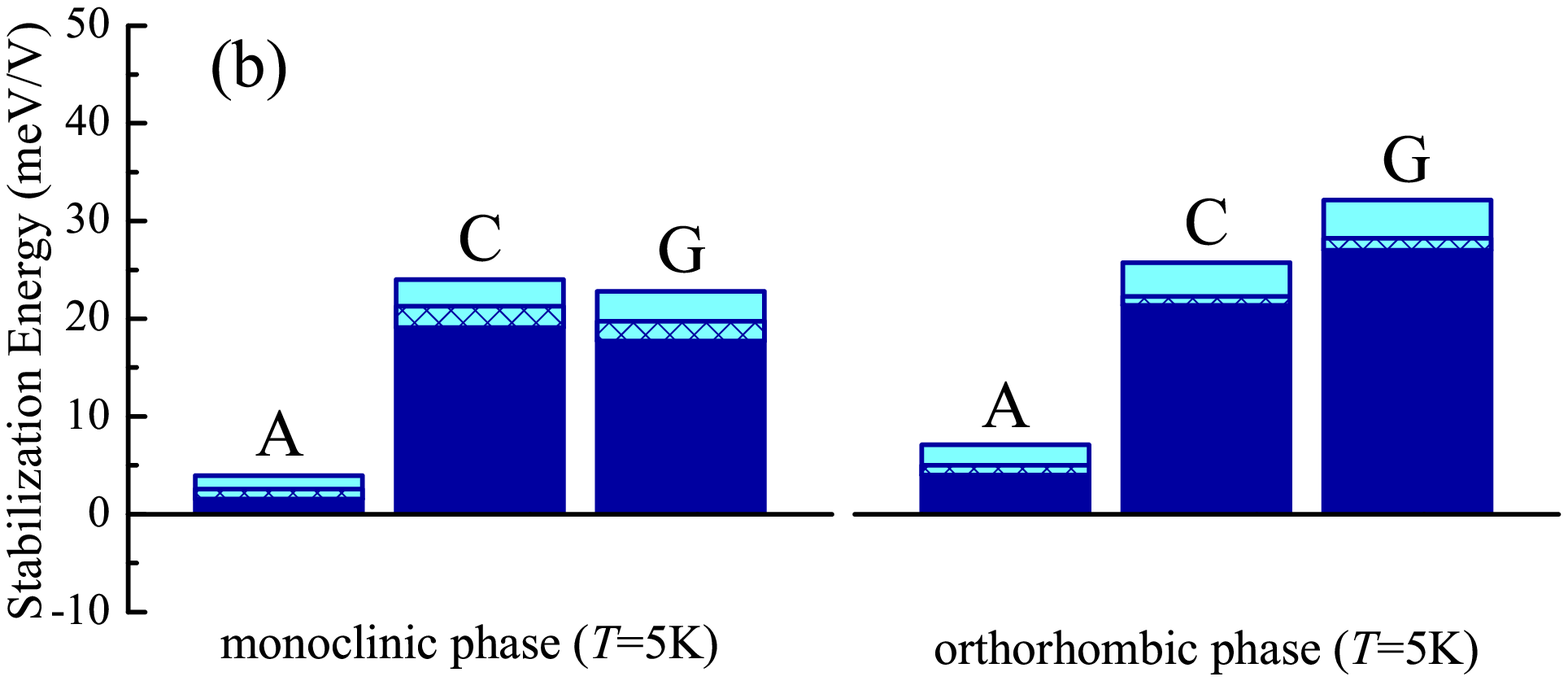}

  \includegraphics[width=0.75\textwidth]{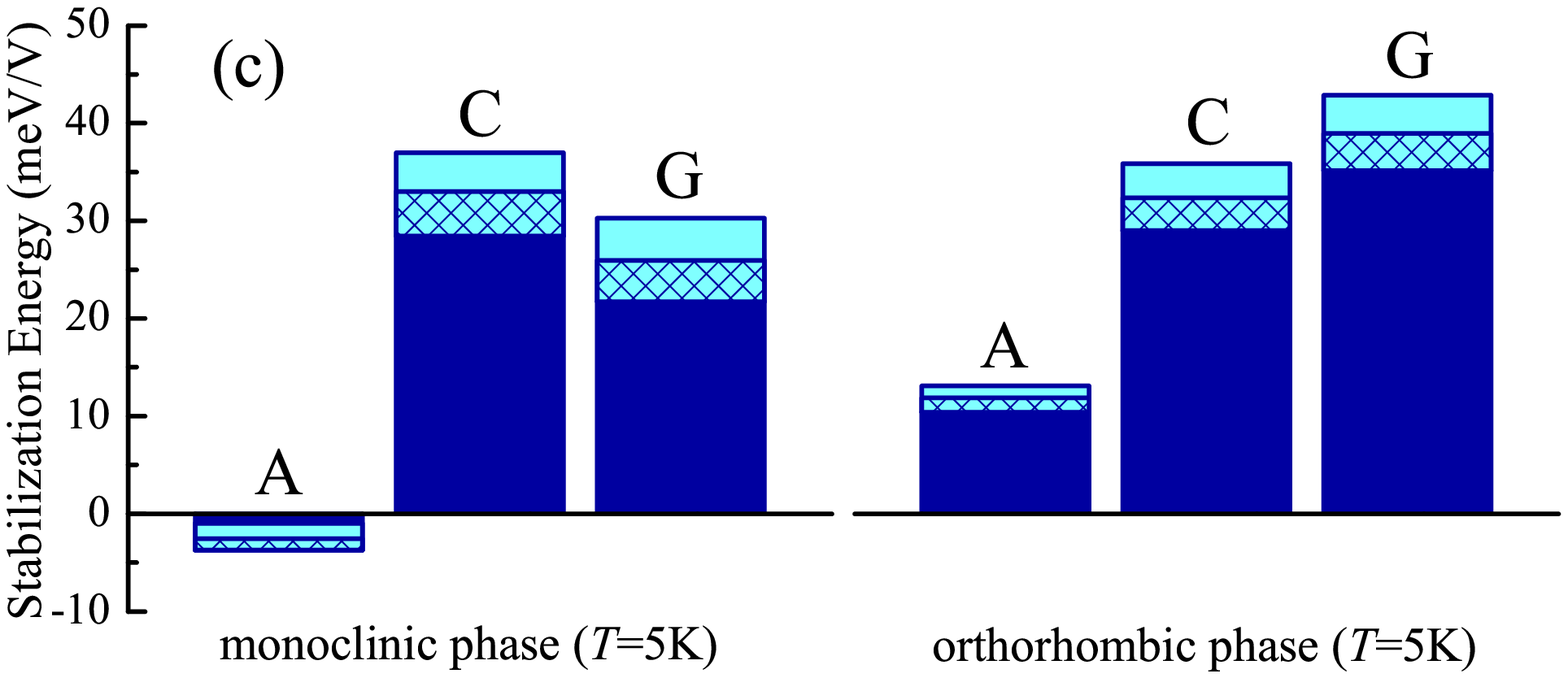}
\caption{Stabilization energies of the main antiferromagnetic states in
SmVO$_3$ (a), GdVO$_3$ (b), and TbVO$_3$ (d)
relative to the ferromagnetic state as obtained in the Hartree-Fock approximation (dark blue area)
and after taking into account the correlation interactions in the second order of
perturbation theory (light blue area) and in the framework of the T-matrix theory (hatched area).
Results for the orthorhombic and monoclinic phases
(obtained using the experimental crystal structure at $T$$=$ 5 K)
are shown correspondingly in the left and right parts
of the figure.}
\label{fig:TotalEnergySmGdTb}       % Give a unique label
\end{figure*}

  Let us consider the next group of compounds $R$VO$_3$ ($R$$=$ Sm, Gd, and Tb),
which below the magnetic transition
temperature exist \textit{simultaneously} in the
monoclinic and orthorhombic phases. The experimental crystal structure parameters,
which have
have been refined for both phases, can be found in \cite{SagePRL,SageThesis}.
In the present study we use the data for $T$$=$ 5 K. The stabilization energies
for several AFM states, calculated for two crystallographic modifications, are
shown in Fig.~\ref{fig:TotalEnergySmGdTb}.
Their behavior
is quite consistent with the expectations based on the simplified CF theory:
the monoclinic structure tends to stabilize the C-type
AFM ordering, while the orthorhombic structure -- G-type AFM ordering,
although with some exceptions. For example, in the orthorhombic phase
of SmVO$_3$, the total energy is lower for the C-type AFM state,
although the next G-type AFM state is very close in energy
(the energy difference between G- and C-type AFM states is only 0.8 meV per one V-site).
It seems that the quasidegeneracy
of the G- and C-type AFM states is the generic feature of SmVO$_3$
and GdVO$_3$ (but not of TbVO$_3$). In the monoclinic phase of GdVO$_3$,
the C-type AFM state has the lowest energy. However, the next G-type AFM state
is only 1.4 meV higher than the C-type AFM state.
This picture seems to be consistent with the experimentally observed phase coexistence,
which means that
there should be several phases competing in a narrow energy range. For example, the
proximity of G- and C-type AFM states in one
of the
crystallographic phases may indicate at the importance of magnetic forces
in driving the transition to another phase.
More specifically, in SmVO$_3$ at $T$$=$ 5 K, the smaller orthorhombic fraction
($\sim$25\% of the volume~\cite{SagePRL})
should be nearly unstable, because the magnetic transition to the C-type AFM state
can cause the structural transition. On the contrary,
in GdVO$_3$ the smaller volume fraction
is monoclinic ($\sim$30\% at $T$$=$ 5 K~\cite{Yusupov}),
which is also nearly unstable, because the change of the crystal structure in this case can be
induced by the
magnetic transition to the G-type AFM state.
Such a transition can be caused either by
the exchange striction
(basically, the change of the lattice parameters associated with the
change of the magnetic state)
or by the change of the orbital
structure, which would further minimize the energy of SE interactions~\cite{KugelKhomskii}.
The relative role of these two mechanisms will be clarified
in the next section.
The quasidegeneracy of magnetic states in the monoclinic phase of GdVO$_3$ may be also
resolved through the additional symmetry-breaking transition and emergence of a new
ordered phase, as it was suggested in some experimental reports~\cite{Yusupov}.

  Similar to the previous group of compounds, the correlation interactions in
$R$VO$_3$ ($R$$=$ Sm, Gd, and Tb) play an important role
and additionally stabilize
the ``correct'' magnetic ground state: C-type AFM for
monoclinic systems and G-type AFM
for orthorhombic ones.

\begin{figure*}
  \includegraphics[width=0.75\textwidth]{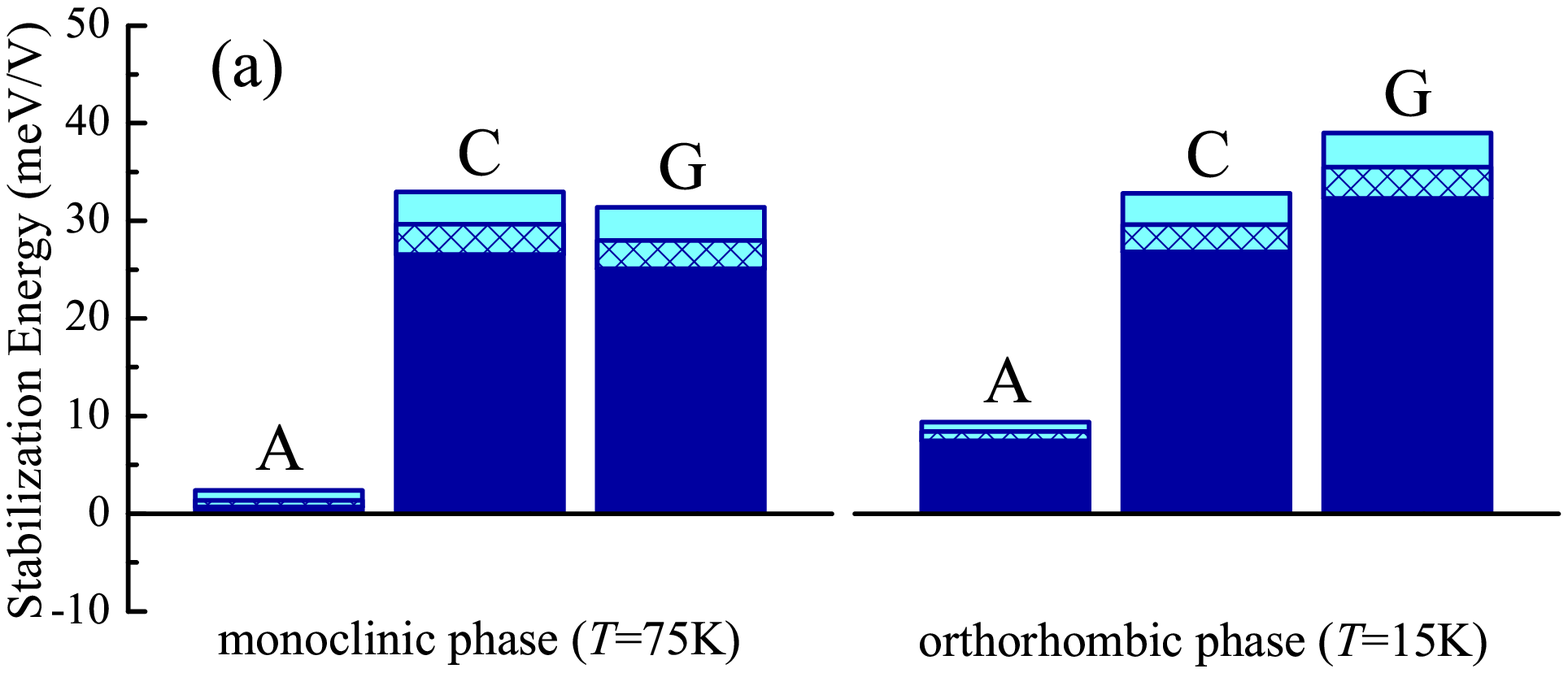}

  \includegraphics[width=0.75\textwidth]{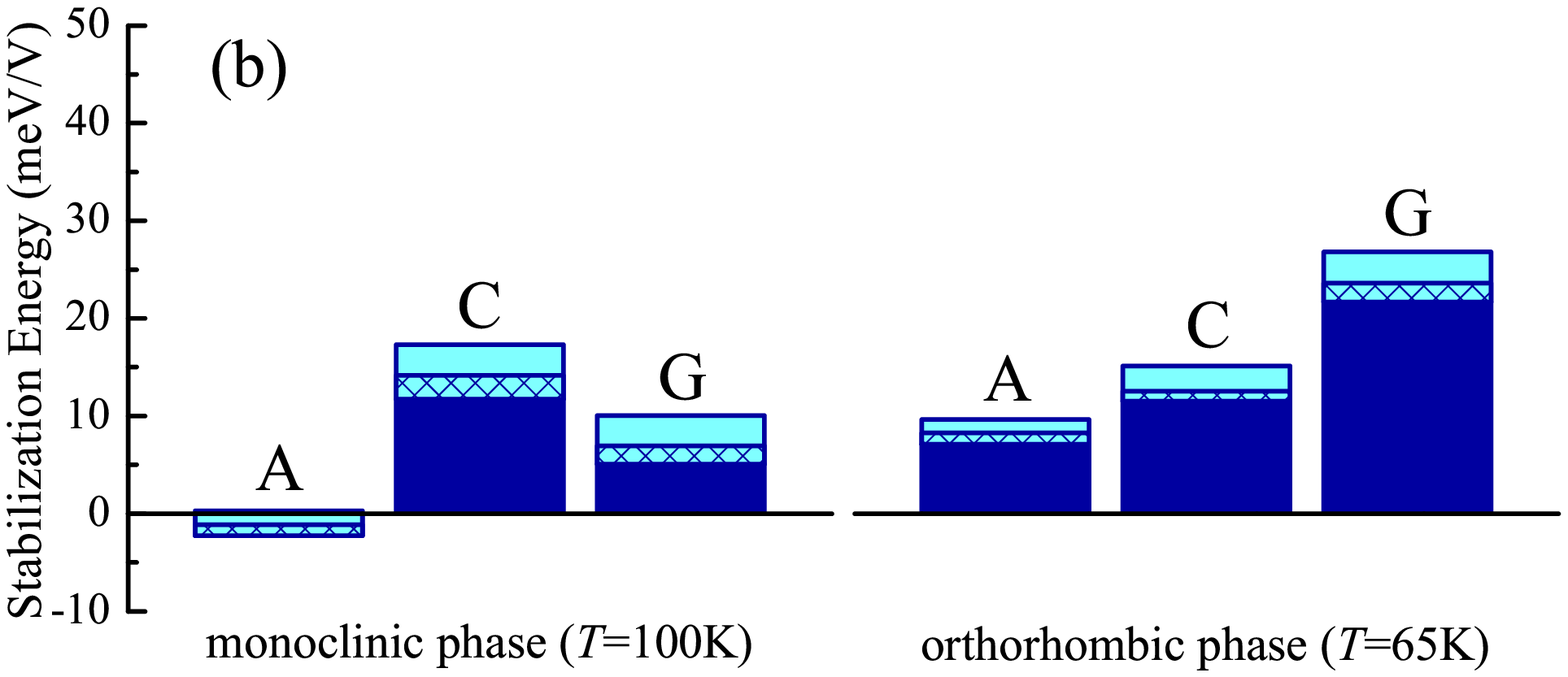}
\caption{Stabilization energies of the main antiferromagnetic states in
YbVO$_3$ (a) and YVO$_3$ (b)
relative to the ferromagnetic state as obtained in the Hartree-Fock approximation (dark blue area)
and after taking into account the correlation interactions in the second order of
perturbation theory (light blue area) and in the framework of the T-matrix theory (hatched area).
Results for the orthorhombic and monoclinic phases
are shown correspondingly in the left and right part
of the figure.}
\label{fig:TotalEnergyYbY}       % Give a unique label
\end{figure*}

  Typical representatives of the last group of compounds are YbVO$_3$ and YVO$_3$.
Similar to the previous case,
below the N\'{e}el temperature,
YbVO$_3$ and YVO$_3$
can be found both in orthorhombic and monoclinic modifications.
However, contrary to $R$VO$_3$ ($R$$=$ Sm, Gd, and Tb), these two phases emerges
\textit{consequently}, in two different temperature intervals.
For examples, below $T_s$$=$ 77 K, YVO$_3$ crystallizes in the orthorhombic modification,
and above $T_s$ -- in the monoclinic one. The construction and solution of the
model Hamiltonian for YVO$_3$ have been discussed in details in \cite{review2008,JETP07,PRB06b}.
Here we summarize the main results for the total energies. They are shown in Fig.~\ref{fig:TotalEnergyYbY}
(together with YbVO$_3$, which exhibits quite a similar behavior).
In this case, the type of the magnetic ground state realized in each temperature range
is controlled
by the crystals structure:
the low-temperature orthorhombic phase coexists with the G-type AFM ordering, while
the high-temperature monoclinic phase coexists with the C-type AFM ordering.
Each magnetic ground state can be reproduced at the level of the HF approximation and
is
additionally stabilized by correlation interactions, considered in the framework of the
second-order perturbation theory as well as in the T-matrix theory.

\subsection{Interatomic magnetic interactions as a probe of the orbital state}
\label{MagneticInteractions}

  In this section we will discuss some limitations of the CF theories.
Since the CF splitting is finite, the orbital degrees of freedom are not
fully quenched and can be affected by other interactions, which generally compete with the
crystal field. Particularly, another major factor ``reshaping'' the orbitals
is related to the superexchange processes~\cite{KugelKhomskii}. In this case,
the occupied orbitals will tend to additionally adjust their form for each magnetic state by
minimizing the energy of SE interactions. The magnitude of this effect
is controlled by the ratio of ${\rm Tr}_L ( \hat{t}_{ij}\hat{t}_{ji} )/{\cal U}$
to the CF splitting.

  Thus, the main question, which will be addressed in this section is
\textit{how flexible are the orbital degrees of freedom in $R$VO$_3$}?
A very useful tool for the analysis of this kind of problem is the
interatomic magnetic interactions (\ref{eqn:JHeisenberg}).
Since interatomic magnetic interactions are defined \textit{locally}, via infinitesimal
rotations of spins, they carry an information about the
details of the orbital state in each magnetic state.
Thus, the parameters of the Heisenberg model (\ref{eqn:HHeisenberg}) depend on the
magnetic state in which they are calculated. However, it is not an artifact
of calculations or the Heisenberg model itself.
This dependence has a clear physical meaning and reflects the changes of the
orbital configuration in each magnetic state.

\begin{figure*}
  \includegraphics[width=0.375\textwidth]{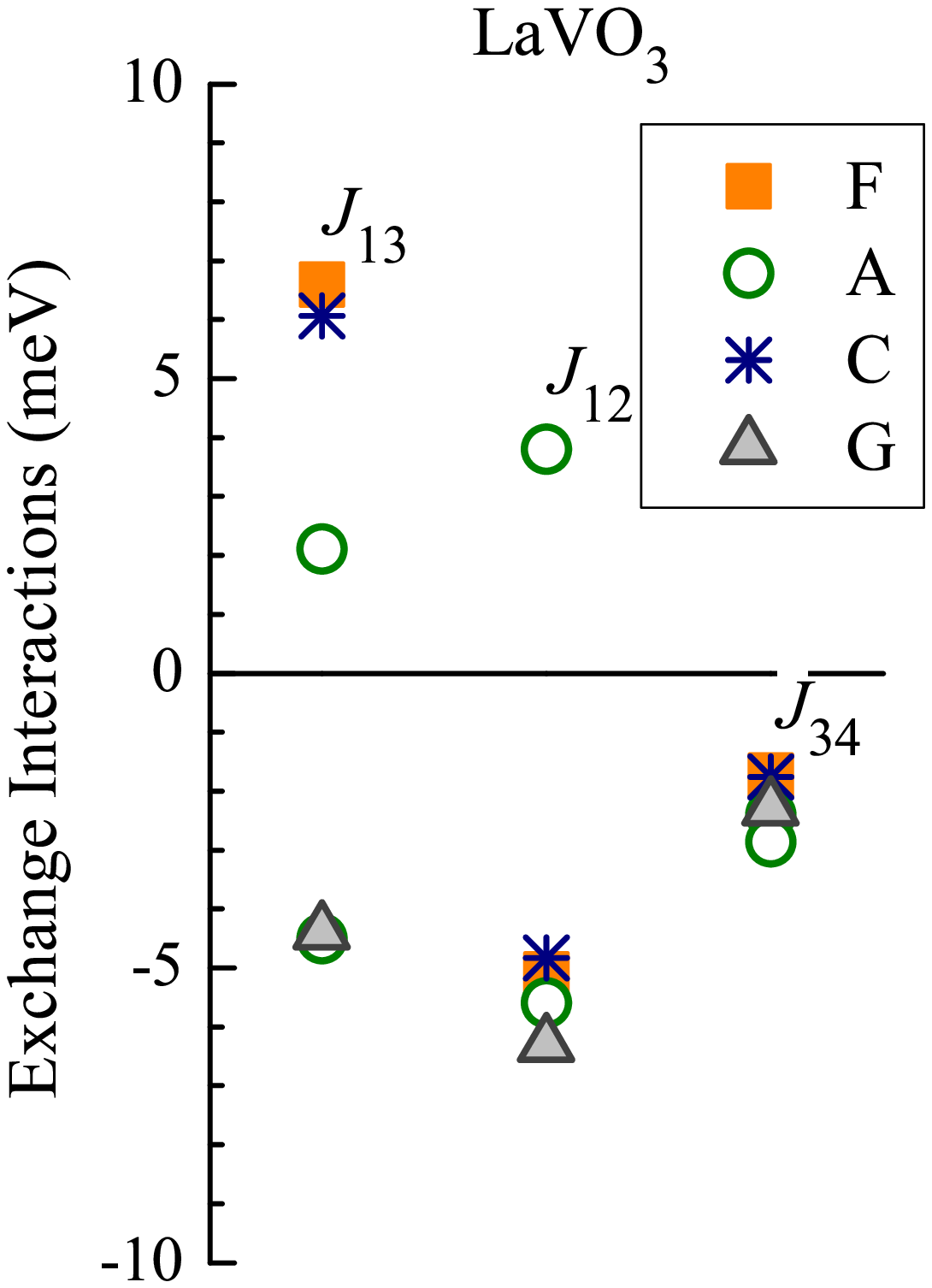}
  \includegraphics[width=0.375\textwidth]{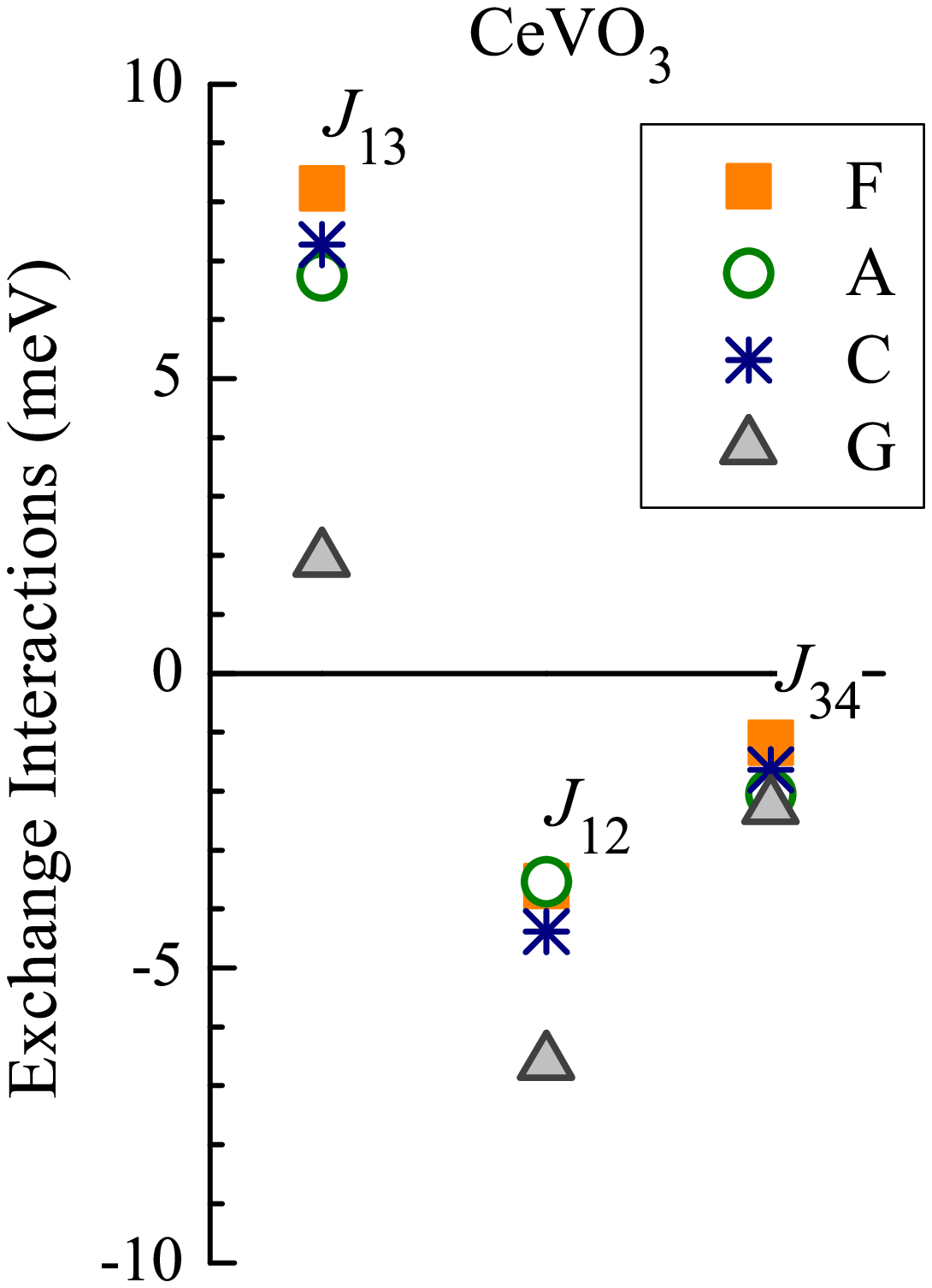}

  \includegraphics[width=0.375\textwidth]{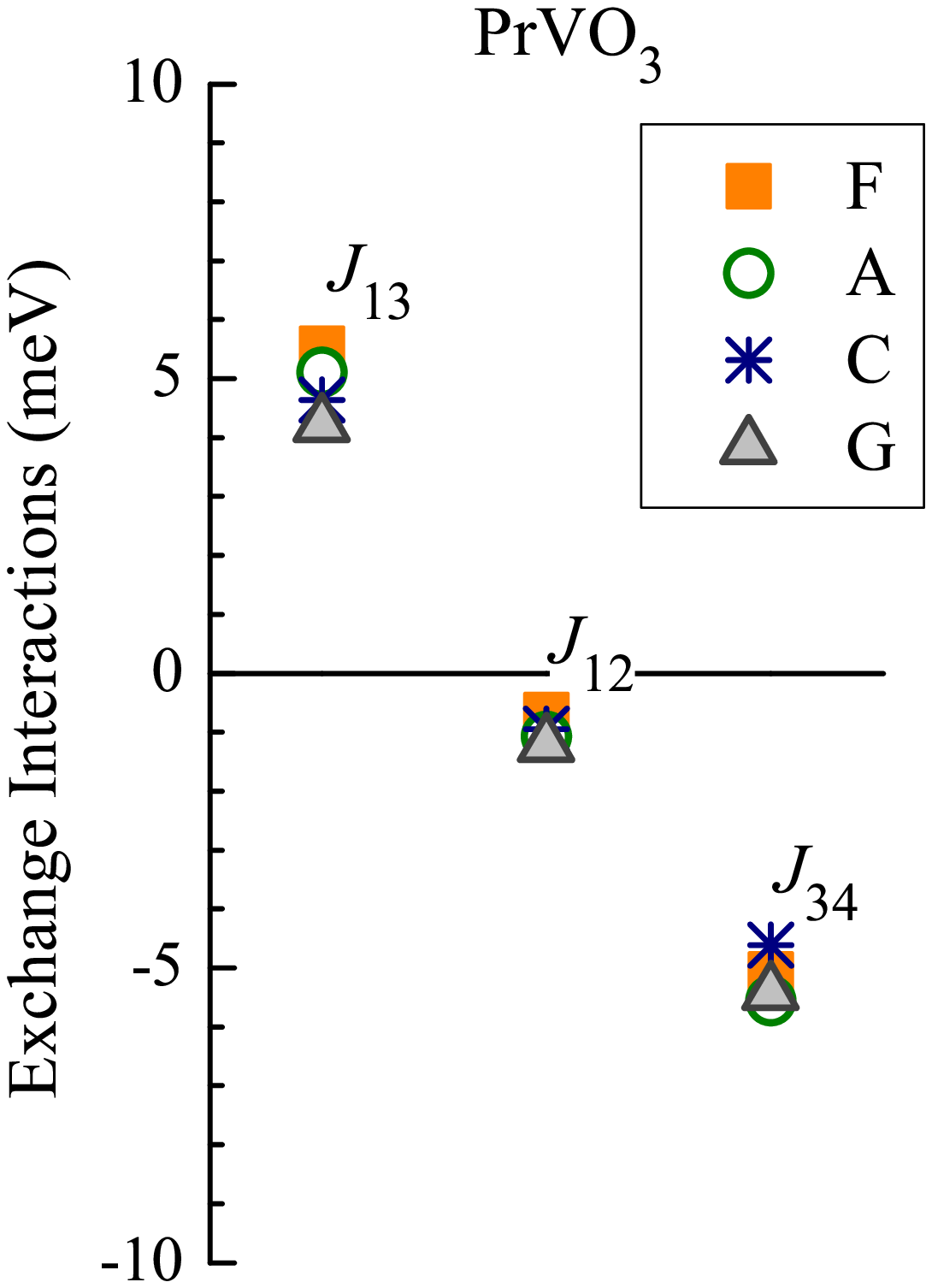}
  \includegraphics[width=0.375\textwidth]{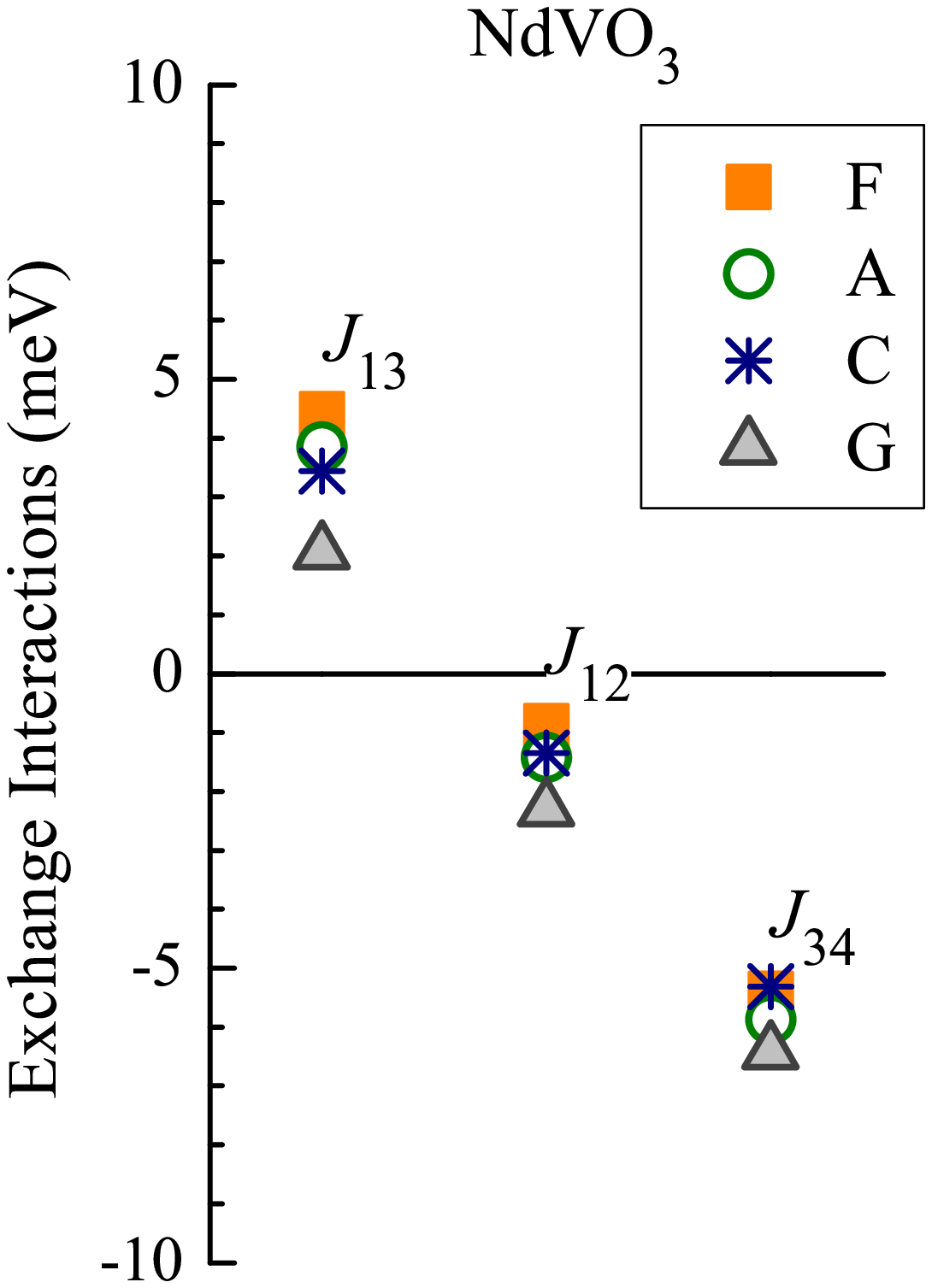}
\caption{Nearest-neighbor magnetic interactions in
LaVO$_3$, CeVO$_3$, PrVO$_3$, and NdVO$_3$
calculated in the ferromagnetic (F) and in the A-, C, and G-type
antiferromagnetic states using the expression (\protect\ref{eqn:JHeisenberg})
for the infinitesimal rotations of spins.}
\label{fig:MagneticInteractionsLaCePrNd}       % Give a unique label
\end{figure*}

\begin{figure*}
  \includegraphics[width=0.75\textwidth]{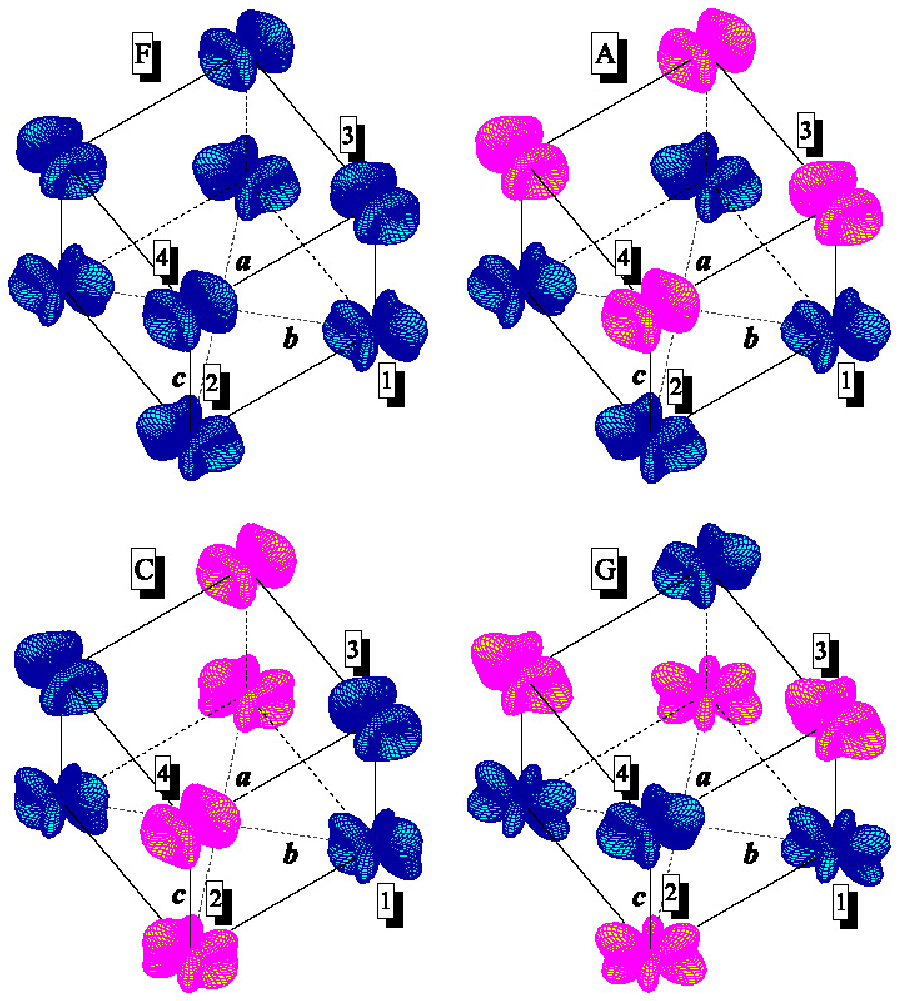}
\caption{Orbital ordering realized in the Hartree-Fock approximation
for the ferromagnetic (F)
and A-, C-, and G-type antiferromagnetic states of CeVO$_3$.
Different magnetic sublattices are shown by different colors.}
\label{fig:OrbitalOrderCe}       % Give a unique label
\end{figure*}

  This effect is clearly seen in the behavior of nearest-neighbor magnetic
interactions in $R$VO$_3$ ($R$$=$ La, Ce, Pr, and Nd), crystallizing in the
monoclinic structure (Fig.~\ref{fig:MagneticInteractionsLaCePrNd}).
Let us consider first the least distorted compound CeVO$_3$ (similar
situation
occurs in LaVO$_3$,
which was considered in \cite{PRB06b}). Distribution of the charge densities
associated with the occupied $t_{2g}$-orbitals (the so-called orbital ordering),
which was obtained by minimizing the HF energy for different magnetic states,
is shown in Fig.~\ref{fig:OrbitalOrderCe}.
This compound has two V-sublattices: ``less distorted''
(according to
the values of the CF-splitting -- Fig.~\ref{fig:CF}) sublattice
(1,2) and ``more distorted'' one (3,4).
Then, one can see even visually that the shape of the orbitals
changes depending on the magnetic state.
As expected,
the orbitals in the ``less distorted'' sublattice (1,2) are more flexible.
The largest change of the occupied orbitals occurs in the G-type AFM state.
All these tendencies are reflected in the behavior of interatomic magnetic
interactions (Fig.~\ref{fig:MagneticInteractionsLaCePrNd}), which tend to additionally
stabilize the magnetic state in which they are calculated.
Particularly, the FM interaction $J_{13}$ decreases drastically
in the G-type AFM state. On the  other hand, the strength of the AFM
interaction $J_{12}$ increases. Similar situation occurs in
LaVO$_3$~\cite{PRB06b}: due to the rearrangement of the occupied orbitals, not only
C-, but also
G-type AFM state becomes locally stable (all nearest-neighbor interactions
are antiferromagnetic in the G-type AFM state, while $J_{12}$ is
ferromagnetic in the C-type AFM state). Nevertheless, the C-type AFM state
has lower energy and is realized as the true magnetic ground
state of LaVO$_3$ (Fig.~\ref{fig:TotalEnergyLaCePrNd}).
Furthermore, the orbital ordering in the A-type AFM state of LaVO$_3$
further breaks the monoclinic symmetry and makes two inequivalent
sublattices in the plane (1,2) (which is reflected in the splitting
of each of the magnetic interaction $J_{12}$ and $J_{13}$ in two types --
ferromagnetic and antiferromagnetic).
When the crystal distortion increases in the direction
LaVO$_3$ $\rightarrow$ CeVO$_3$ $\rightarrow$ PrVO$_3$ $\rightarrow$ NdVO$_3$
(see Fig.~\ref{fig:CF}), the orbitals become less flexible, and the
magnetic interactions in PrVO$_3$ and NdVO$_3$ only weakly depend
on the magnetic state in which they are calculated. Thus, in the last two compounds, we have
more or less conventional CF scenario of the orbital ordering and
related to it interatomic magnetic interactions.

\begin{figure*}
  \includegraphics[width=0.5\textwidth]{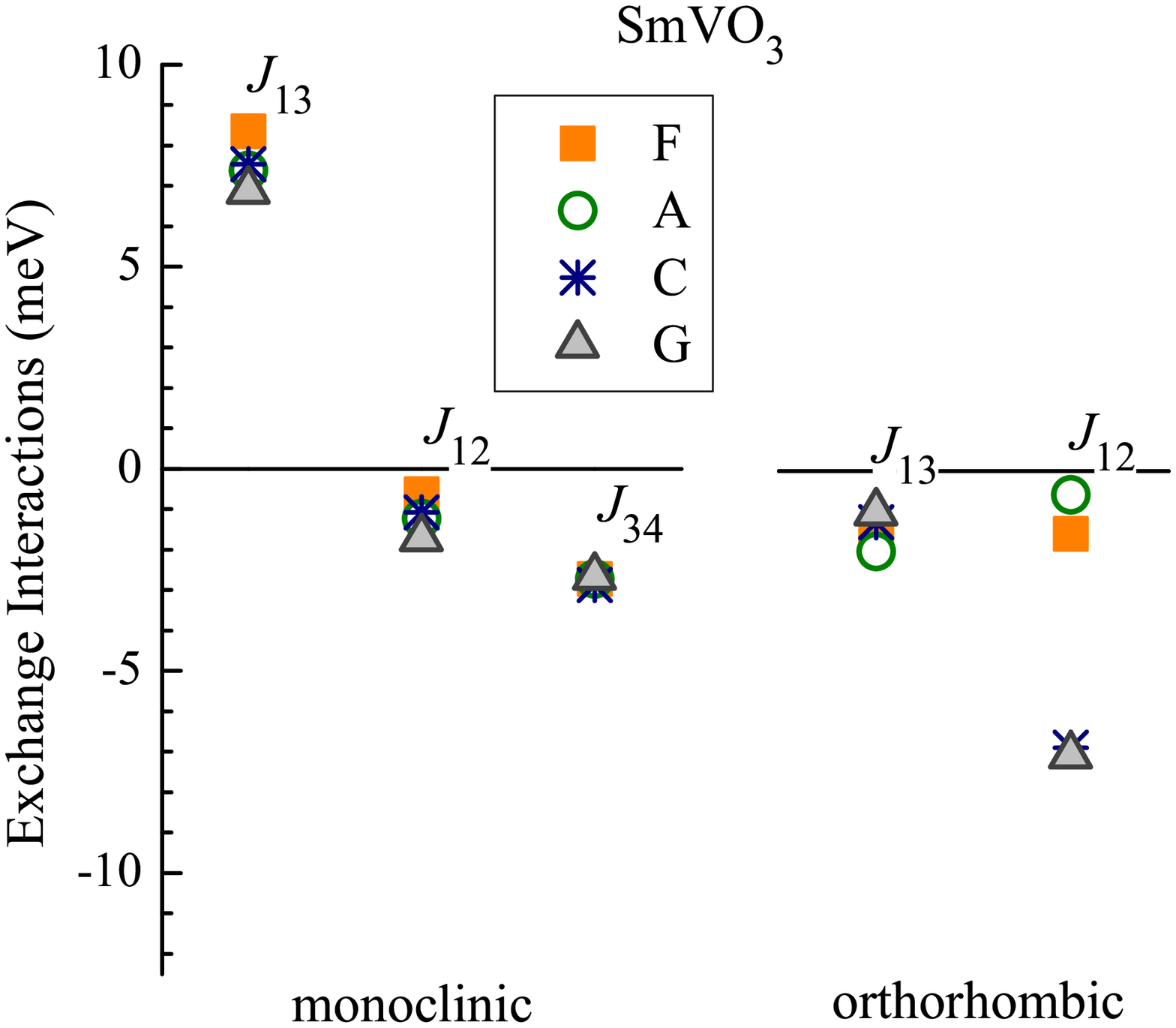}

  \includegraphics[width=0.5\textwidth]{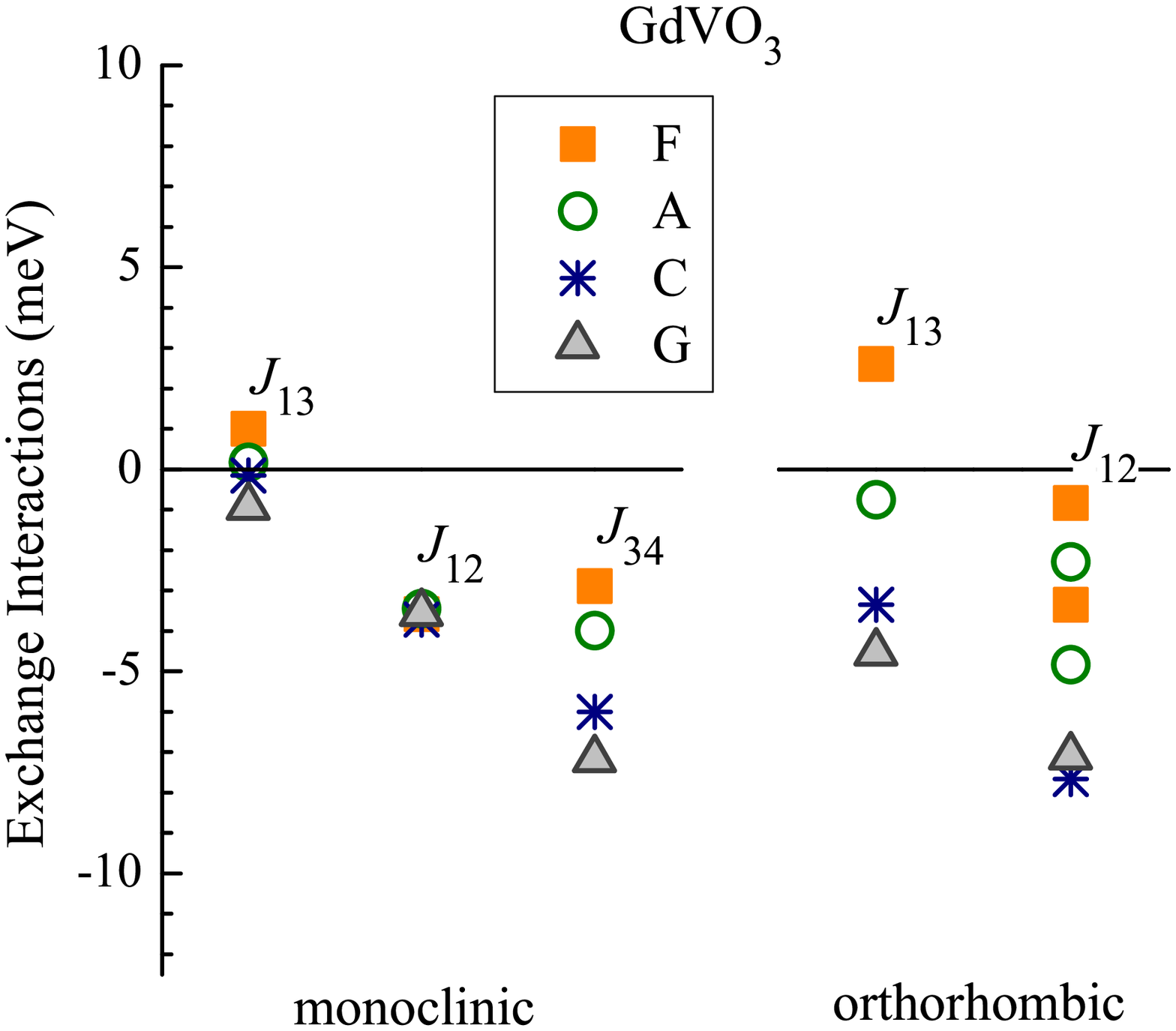}

  \includegraphics[width=0.5\textwidth]{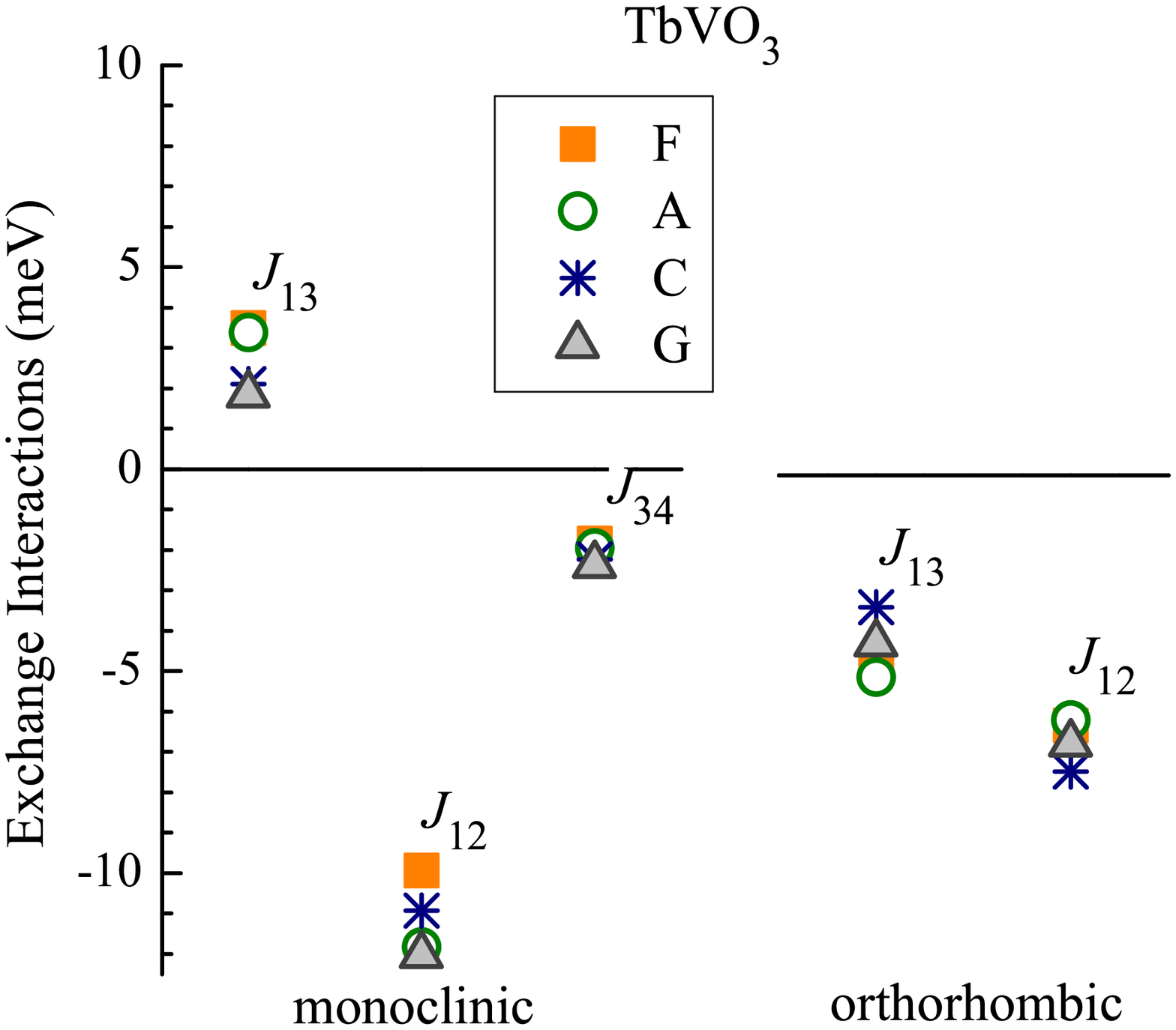}
\caption{Nearest-neighbor magnetic interactions in
the monoclinic (left) and orthorhombic (right) phases of
SmVO$_3$, GdVO$_3$, and TbVO$_3$,
calculated in the ferromagnetic (F) and in the A-, C, and G-type
antiferromagnetic states using the expression (\protect\ref{eqn:JHeisenberg})
for the infinitesimal rotations of spins.}
\label{fig:MagneticInteractionsSmGdTb}       % Give a unique label
\end{figure*}

\begin{figure*}
  \includegraphics[width=0.75\textwidth]{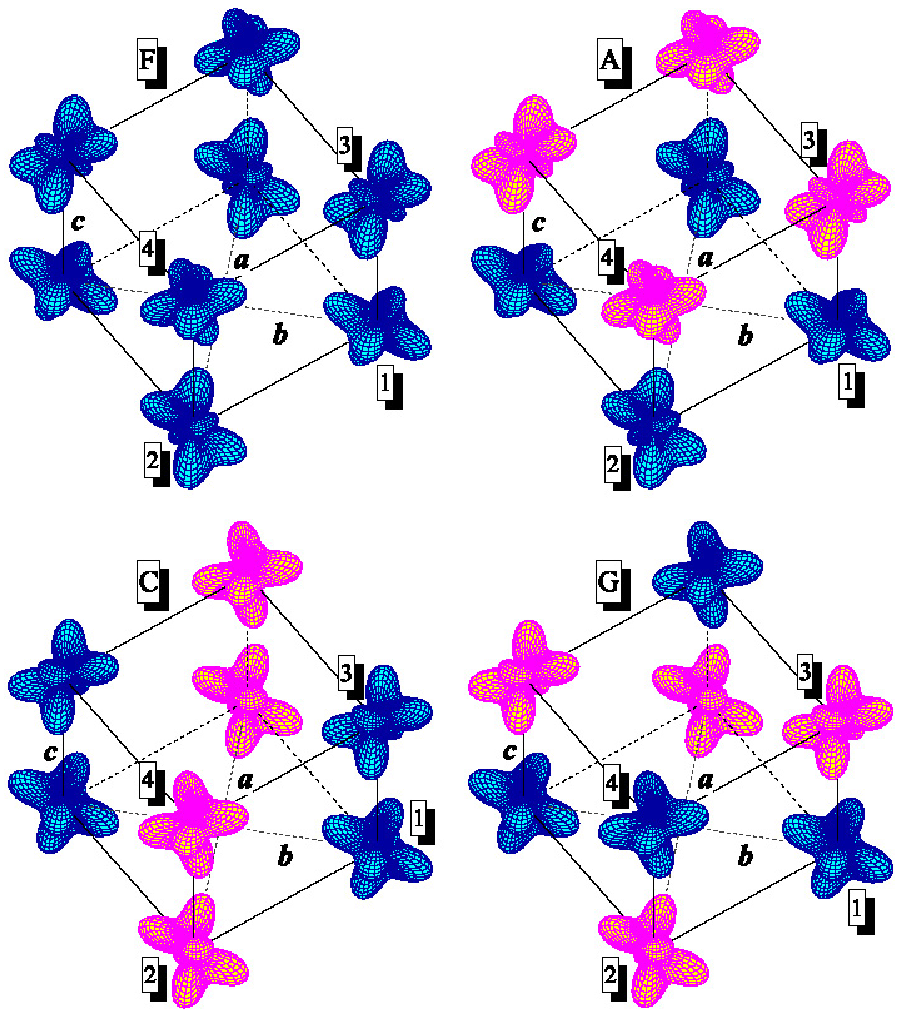}
\caption{Orbital ordering realized in the Hartree-Fock approximation
for the ferromagnetic (F)
and A-, C-, and G-type antiferromagnetic states of
orthorhombic SmVO$_3$ (at $T$$=$ 5 K).
Different magnetic sublattices are shown by different colors.}
\label{fig:OrbitalOrderSm}       % Give a unique label
\end{figure*}

  The behavior of interatomic magnetic interactions in $R$VO$_3$ ($R$$=$ Sm, Gd, and Tb),
crystallizing simultaneously in the monoclinic and orthorhombic structures, is
explained in Fig.~\ref{fig:MagneticInteractionsSmGdTb}.
As was pointed out in the previous section, one of the interesting features
in this regime is the quasidegeneracy of the C- and G-type AFM states, realized in
the orthorhombic phase of SmVO$_3$ and in the monoclinic phase of GdVO$_3$.
Therefore, we have to address the question whether this quasidegeneracy is related to
the reconstruction of the orbitals ordering, which will further affect the
interatomic magnetic interactions and make them specific for each
magnetic state, or to some other effects, such as the
exchange striction.
The typical example of the orbital ordering
realized in the orthorhombic phase of SmVO$_3$ is shown in Fig.~\ref{fig:OrbitalOrderSm}.
The orbital degrees of freedom in these compounds are
indeed rather flexible and to some extent are able to adjust the change of the magnetic state.
Nevertheless, one important aspect is that the orbital ordering does not significantly change
between the C- and G-type AFM states. In fact, the orbital ordering pattern in the orthorhombic phase of SmVO$_3$
can be of two types: one type is realized for the C- and G-type AFM states and
another type -- for the F-state and the A-type AFM state. The orbital ordering within each group of states is practically identical.
This behavior of the orbital degrees of freedom is reflected in the behavior of
interatomic magnetic interactions (Fig.~\ref{fig:MagneticInteractionsSmGdTb}): similar to the orbital ordering,
there are two sets of the parameters $\{ J_{ij} \}$, one of which acts in the F- and A-states and another one --
in the C- and G-states. The difference is mainly reflected in the behavior of in-plane interaction $J_{12}$.
Similar situation occurs in the monoclinic phase of GdVO$_3$, where two sets of the in-plane
interactions $J_{34}$ are associated with different orbital states, realized correspondingly for the FM and
A-type AFM ordering and for the C- and G-type AFM ordering.
Thus, the quasidegeneracy of the C- and G-type AFM states in SmVO$_3$ and GdVO$_3$ cannot be
related to the change of the orbital state. A more likely scenario is the exchange striction,
where the change of the lattice parameters can easily change the sign of the weak inter-plane interaction $J_{13}$.

\begin{figure*}
  \includegraphics[width=0.5\textwidth]{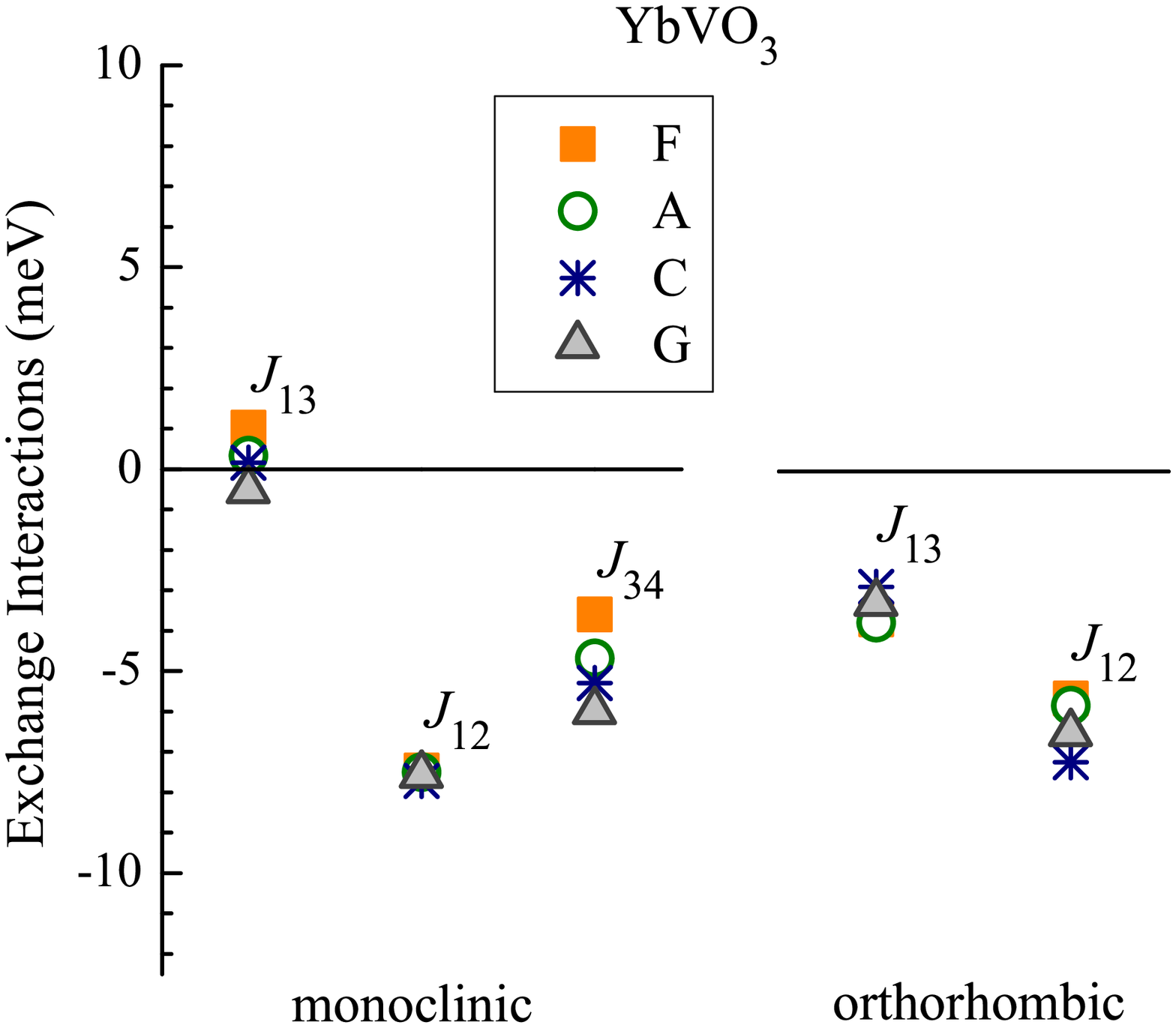}

  \includegraphics[width=0.5\textwidth]{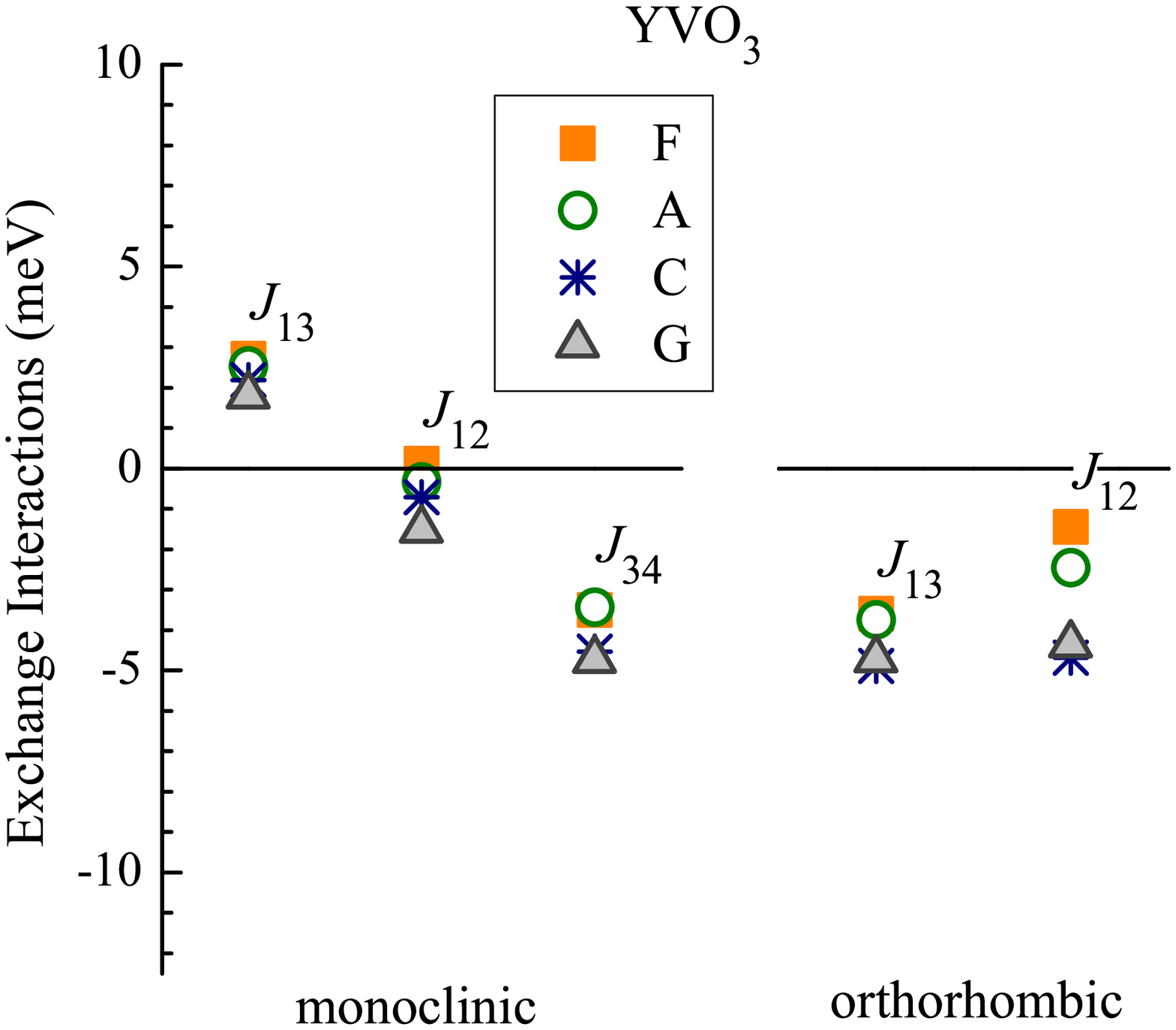}
\caption{Nearest-neighbor magnetic interactions in
the monoclinic (left) and orthorhombic (right) phases of
YbVO$_3$ and YVO$_3$,
calculated in the ferromagnetic (F) and in the A-, C, and G-type
antiferromagnetic states using the expression (\protect\ref{eqn:JHeisenberg})
for the infinitesimal rotations of spins.}
\label{fig:MagneticInteractionsYbY}       % Give a unique label
\end{figure*}

  The interatomic magnetic interactions for the last two compounds, YbVO$_3$ and YVO$_3$,
which exhibits the consecutive monoclinic-to-orthorhombic transition with the decrease of
temperature are shown in Fig.~\ref{fig:MagneticInteractionsYbY}.
Details of the orbital ordering and interatomic magnetic interactions for YVO$_3$ can be
found in~\cite{PRB06b}.
In this case, the orbitals degrees of freedom can remain flexible and the
interatomic magnetic interactions depend on the magnetic states (especially
for the orthorhombic phase of YVO$_3$).
Nevertheless, the effect is not so strong as in SmVO$_3$ and GdVO$_3$.
Moreover, the form of the orbital ordering and interatomic magnetic interactions,
obtained for the C- and G-type AFM states in each of the crystallographic modification,
is practically the same (similar to SmVO$_3$ and GdVO$_3$). Thus, the orbital fluctuations
do not seem to be the trigger for the C-G-type AFM transition and the concomitant
monoclinic-to-orthorhombic phase transition. The present results are more consistent with
the scenario where the structural transition goes first, while the spin
and orbital degrees of freedom mainly follow the change of the crystal structure.

\section{Summary and Conclusions}
\label{summary}

  We have reviewed the main ideas of realistic modeling of the strongly correlated systems.
This is a new direction of the electronic structure calculations, which is especially suit
for complex oxides materials. In the present context, the term ``complex'' means the computational
complexity
(the necessity to deal with highly distorted crystal structures
having very low symmetry and many atoms in the unit cell) as well as the methodological
complexity
(the necessity to go beyond conventional approximations in the
electronic structure calculations, such as the local-density approximation).
The basic idea of realistic modeling is to construct an effective low-energy model
for the states close to the Fermi level, derive all the parameters rigorously, on
the basis of first-principles electronic structure calculations, and to solve this model
using modern many-body techniques.
Thus, realistic modeling combines the accuracy and predictable power of first-principles electronic
structure calculations with the flexibility and insights of the model analysis.

  It is certainly true that approximations are inevitable in such an approach
(and the form of the low-energy model itself is the main approximation!).
However, we would like to emphasize again that apart from these approximations,
the entire procedure is parameter-free.
Namely, we do not need anymore to deal with
numerous adjustable parameters.
Instead, the state of the discussion is brought to a qualitatively
new level: how to improve the approximations used for the definition and calculation
of the model parameters.

  The abilities of this method were demonstrated on the wide series of vanadates
$R$VO$_3$ ($R$$=$ La, Ce, Pr, Nd, Sm, Gd, Tb, Yb, and Y)
with distorted perovskite structure. Particular attention was paid
to computational tools, which can be used for the microscopic analysis
of different spin
and orbital states
realized
in the partially filled $t_{2g}$-band. Meanwhile, we were able to solve a number of
fundamental and materialogical problems related to the origin of
the C- and G-type AFM states in these compounds.
The first applications of realistic modeling are very encouraging.
We hope that these ideas will continue to develop in the future to become a powerful tool for
theoretical analysis of complex oxide materials and other strongly correlated systems.

\begin{acknowledgements}
I am grateful to G.R. Blake for providing details of the experimental
crystal structure
for the series of compounds $R$VO$_3$. I also wish to thank the support
from the Federal Agency for Science and Innovations, grant No. 02.740.11.0217,
during my stay at the Ural State Technical University (Ekaterinburg, Russia).
\end{acknowledgements}



\end{document}